\documentclass[twocolumn,secnumarabic,amssymb, nobibnotes, aps, prd]{revtex4-1}

\usepackage{wrapfig}
\usepackage{dsfont}
\usepackage{amscd}
\usepackage{amsmath}
\usepackage{amssymb}
\usepackage{graphicx}
\usepackage{bm}
\usepackage{newtxtext}
\usepackage{booktabs,caption} 

\usepackage{comment}
\usepackage[tikz]{bclogo}
\presetkeys{bclogo}{
ombre=true,
epBord=1,
couleur = blue!15!white,
couleurBord = red,
arrondi = 0.2,
logo=\bctrombone
}{}

\usepackage{subcaption}
\usepackage{sidecap}
\usepackage{grffile}
\usepackage{xcolor}
\usepackage{scalerel}

\definecolor{rred}{rgb}{0.7,0,0.1}
\definecolor{greenrb}{rgb}{0.2,0.6,0.2}

\newcommand{\mk}{\color{black}}

\newcommand{\mkr}{\color{black}}
\newcommand{\mkd}{\color{black}}

\newcommand{\mkdd}{\color{black}}

\def\bi{\begin{itemize}}
\def\ei{\end{itemize}}

\def\bea{\begin{equation} \begin{aligned}}
\def\eea{\end{aligned} \end{equation}}
\def\beas{\begin{equation*} \begin{aligned}}
\def\eeas{\end{aligned} \end{equation*}}
\def\bes{\begin{equation*}}
\def\ees{\end{equation*}}
\def\d{\, \mathrm{d}}
\def\be{\begin{equation}}
\def\ee{\end{equation}}

\renewcommand\d{\, \mathrm{d}}

\newcommand{\ellL}{\mathcal{L}}

\def\u{\boldsymbol{u}}
\def\vb{\boldsymbol{v}}
\def\U{\boldsymbol{U}}
\def\x{\boldsymbol{x}}
\def\y{\boldsymbol{y}}

\def\z{\boldsymbol{z}}

\def\Fb{\boldsymbol{F}}
\def\Sigmab{\boldsymbol{\Sigma}}
\def\Cb{\boldsymbol{C}}

\def\e{{\color{black} \epsilon}}
\def\F{{\color{black}{\mathcal{F}}}}

\def\W{{\boldsymbol{W}_t}}

\def\M{\mathcal M}  
\def\m{\mathfrak m}

\newcommand{\R}{\mathbb{R}}

\newcommand{\GGG}{\mathcal{G}}

\newcommand{\LLL}{\mathcal{L}}

\newcommand{\xx}{\mathbf{x}}

\newcommand{\FF}{\mathbf{F}}
\newcommand{\GG}{\mathbf{G}}

\newcommand{\dd}{\mathrm{d}}

\newcommand{\lp}{\left(}
\newcommand{\rp}{\right)}

\newcommand{\beq}{\begin{equation}}
\newcommand{\eeq}{\end{equation}}
\newcommand{\beqs}{\begin{subequations}}
\newcommand{\eeqs}{\end{subequations}}
\newcommand{\balign}{\begin{align}}
\newcommand{\ealign}{\end{align}}

\begin{document}

\title{Theoretical tools for understanding the climate crisis from Hasselmann's program and beyond}    
 \author{Valerio Lucarini}
\email{Corresponding author. Email address: \texttt{v.lucarini@reading.ac.uk}}\affiliation{Department of Mathematics and Statistics, University of Reading, Reading, RG6 6AX, UK}
 \affiliation{Centre for the Mathematics of Planet Earth, University of Reading, Reading, RG6 6AX, UK}
 
  \author{Micka\"el D. Chekroun}
\email{Email address: \texttt{mchekroun@atmos.ucla.edu}} \affiliation{Department of Earth and Planetary Sciences, Weizmann Institute of Science, Rehovot 76100, Israel,}
 \affiliation{Department of Atmospheric and Oceanic Sciences, University of California, Los Angeles, CA 90095-1565, USA}

\date{\today}

\begin{abstract}
Klaus Hasselmann's revolutionary intuition {\mkr in climate science} was to take advantage of the stochasticity associated with fast weather processes to probe the slow dynamics of the climate system. This has led to fundamentally new ways to study the response of climate models to perturbations, and to perform detection and attribution for climate change signals. Hasselmann's program has been extremely influential in climate science and beyond. We first summarise the main aspects of such a program using modern concepts and tools of statistical physics and applied mathematics. We then provide an overview of some promising scientific perspectives that might better clarify the science behind the climate crisis and that stem from Hasselmann's ideas. We show how to perform rigorous model reduction by constructing parametrizations in systems that do not necessarily feature a time-scale separation between unresolved and resolved processes. We propose a general framework for explaining the relationship between climate variability and climate change, and for  performing climate change projections. This leads us seamlessly to explain some key general aspects of climatic tipping points. Finally, we show that response theory provides a solid framework supporting optimal fingerprinting methods for detection and attribution.
\end{abstract}

\maketitle    

\section{Introduction}
\begin{figure}
\includegraphics[width=0.45\textwidth]{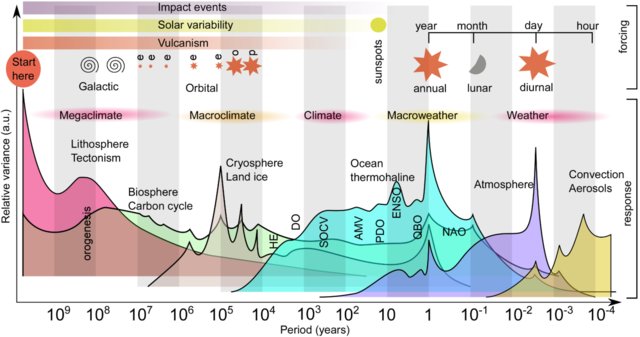}
\caption{\small A Mitchell's diagram \cite{Mitchell1976}  depicting a qualitative representation of the climate variability across a vast range of scales, with indication of the relative role of each climatic subcomponent, and indication, on top, of the acting forcings. From \cite{vonderHeydt2021}.}\label{Fig_Anna}
\end{figure}

The climate is a complex and complicated system comprising of five subsystems - the atmosphere, the hydrosphere, the cryosphere, the biosphere, and the land surface - which differ for physical-chemical features, dominants dynamical processes, and characteristic time scales. Such subsystems are coupled through a complex array of processes of exchange of mass, momentum, and energy \cite{Peixoto1992,Lucarini.ea.2014}.  The climate system is multiscale as it features variability over a vast range of scales, as a result of the interplay of a very diverse array of forcings, instabilities, and feedbacks \cite{Mitchell1976,Ghil.2019}, with different subsystems playing the dominant role in different considered temporal (and spatial) ranges \cite{vonderHeydt2021}, as shown in Fig. \ref{Fig_Anna}. {\color{black}Hence, the natural history of our planet is characterized by continuous changing conditions, with an interplay of rapid, irreversible transitions and of gradual variations \cite{Rothman2017}, with alternating dominance of positive and negative feedbacks depending on the time scale of interest \cite{Arnscheidt2022Feedbacks}.} Major knowledge gaps on the climate system comes from the lack of homogeneous, high-resolution, and coherent observations, because of a) the sheer size and practical {\color{black}in}accessibility of the climate subdomains, b) changes in the technology of data collection in the industrial era, and c) the need to resort to proxy (hence, indirect) data for the pre-industrial epoch. Thus, it is extremely challenging to construct satisfactory theories of climate dynamics and is virtually impossible to develop numerical models able to describe accurately climatic processes over all scales. Typically, different classes of models and different phenomenological theories have been and are still being developed by focusing on specific scales of motion and specific processes \cite{GhilLucarini2020}. 

As a result of the presence of unsteady external forcings and of a very nontrivial internal multiscale dynamics, and of the fact that we experience only one realization of the state of the climate, it is hard to clearly separate climate variability from any climate change signal \cite{Ghil2015}. As documented in successive assessment reports of the Intergovernmental Panel for Climate Change the scientific community has painstakingly come to an agreement regarding 
1) the presence of a statistically significant climate change signal 
with respect to the conditions prevailing in the XIX century {\color{black}(\textit{climate change is real)}; and 2) the possibility of attributing such a signal to anthropogenic causes (\textit{humans are responsible for it}). 

The attribution of the climate change signal requires being able to make a stringent case for a causal link between acting forcings (e.g. land-use change; change in the atmospheric compositions due to human activities; effect of vulcanism; solar variability) and the observed climatic response. According to the Pearl's causality framework \cite{Pearl2009}, this  requires, in turn, comparing the observed state of the climate system with the counterfactual realities where the acting forcings are selectively switched off. Clearly, we do not have access to  counterfactual realities. Instead, we can create reasonable approximations of them by performing numerical simulations with climate models with suitably defined protocols for the forcings. Hence, making statements on climate change attribution takes into account the unavoidable uncertainty due to natural variability and model error \cite{Hasselmann1993b,Hegerl1996,Hasselmann1997,Bindoff2013,IPCC2022}.}

In the course of the years, the research focus has progressively shifted from making statements on globally averaged climatic quantities, to assessing  climate change at regional level, to studying the changes in the higher statistical moments and in extreme events, and to investigating how the climate change can manifest itself in the form of critical transitions. The current focus on the study of extremes \cite{IPCC12} and critical phenomena  (often referred to as tipping points \cite{Lenton.tip.08,Ashwin2012}) has led the scientific community to use expressions like \textit{climate crisis} or \textit{climate emergency} instead of \textit{climate change} \cite{Ripple2021}

\subsection{A Brief Summary of Hasselmann's Program}
In the latter part of the XX century,  Hasselmann proposed a coherent scientific angle on the climate system, with the goal of 
understanding 
climate variability, of detecting and interpreting the climate change signal, and of characterizing the  behaviour of climate models. 
A very informative summary of the so-called Hasselmann program, of some of its key developments in climate science, and of some of its implications for mathematics and physics at large is given in \cite{Imkeller2001,vonStorch2022}. {\color{black}Useful accounts of Hasselmann's work in occasion of him being awarded the 2021 Nobel prize in physics can be found in \cite{Gupta2022,Hegerl2022}.} We  discuss  below three main axes of Hasselmann's program. 

\subsubsection{Stochastic Climate Models}\label{stochasticcm}
The  starting point of this journey  {\mk comes with the seminal work} \cite{Hasselmann1976}, where Hasselmann  proposed {\mk to improve of  modeling of slow climate variables by parameterizing the influence of fast  weather variables by means of an appropriate stochastic forcing.} 
{\mk At the core of this Hasselmann's stochastic program, lies thus the derivation of an} effective model with improved capability to capture the dynamics and the statistical properties of the slow variables. 
Following \cite{Arnold2001}, {\mk we assume that our climate model writes as the following large-dimensional system of ordinary differential equations}
\bea\label{fastslow}
&\dot{\x}=f(\x,\y), \;\; \x \; \textrm{slow climate variables},\\ 
&\dot{\y}=\frac{1}{\delta}g(\x,\y), \;\;\y \; \textrm{fast weather variables},
\eea
where $\x$ (resp.~$\y$) lies in $\mathbb{R}^{d}$ (resp.~$\mathbb{R}^{D}$), and $f$ (resp.~$g$) is a smooth mapping from   
$\mathbb{R}^{d+D}$ into $\mathbb{R}^{d}$ (resp.~$\mathbb{R}^{D}$).

The parameter $\delta$ is aimed at {\mkr controlling the degree of timescale separation between the dynamics of the two sets of variables, and one} typically assumes $0<\delta\ll1$. 
{\mk In mathematical language, the Hasselmann's program consists of deriving in the $\delta\rightarrow 0$ limit an effective reduced equation of Eq.~\eqref{fastslow} to approximate the statistical behaviour of the climate variables.  
When $\y$ follows some fast chaotic dynamics {\mkr as commonly assumed \cite{Kelly2017,cotter2017stochastic}}, the latter is known to take the form of the following   
system of It\^o stochastic differential equations (SDEs),} 
\begin{equation}\label{Hass1}
\d \x= \FF(\x)\d t +\Sigmab(\x)\d \W.
\end{equation}
{\color{black}In the case of climate dynamics, the chaoticity of the dynamics of the $\y$ variables is usually assumed to derive from the fast fluid dynamical instabilities occurring in the atmosphere and in the ocean \cite{GhilChildress1987,Ghil.2019,GhilLucarini2020}. Such instabilities are usually associated with conversion of energy between different forms (e.g. potential to kinetic) and between spatially symmetric and eddy components \cite{Lor67,Peixoto1992}.} {\mk The derivation of such a limiting SDE in the presence of infinite timescale separation has a long history \cite{beck1990brownian, just2001stochastic,majda2001mathematical,Pavliotis2008,Gottwald2013} and may be obtained through diverse routes  pertaining to homogenization \cite{Pavliotis2008}, averaging \cite{khas1963principle}, or singular perturbation techniques \cite{kurtz1973limit,Papanicolaou1974}. The MTV approach \cite{majda2001mathematical,majda2003systematic} building up on such techniques provides a modern treatment of the topic in the context of climate dynamics, including rigorous results when nonlinear self-interactions between e.g.~the fast waves can be modeled by means of Ornstein-Uhlenbeck  processes.}  

{\mk The terms $\FF$ and $\Sigmab$ in Eq.~\eqref{Hass1}  have then intuitive interpretations.  
Typically, the deterministic component $F$ (also called drift term) {\color{black} includes the average contribution of $f(\x,\y)$ to the dynamics of the slow variables, after averaging out the fast variables \cite{just2001stochastic}  {\mkr but potentially may also include other correction terms};  see \cite{Kelly2017} for a detailed analysis.}
The second term, is aimed at parameterizing the effects of the fluctuations left over due to averaging, and takes the form of a state-dependent noise, in which the $d\times p$-matrix $\Sigma$ with (possible) nonlinear entries in $\x$, is driven a vector of increments $\d \W$ (white noise) of a $p$-dimensional Wiener process $ \W$.}

{\color{black}A key take-home message from  Eq.~\eqref{Hass1} is that the impact of unresolved scales of motion on the scales of interest cannot be   reproduced by using bulk formulas (contributing to the drift term) only. This has extremely important practical relevamce: it is the fundamental motivation behind the development of stochastic parametrizations for weather and climate models, some of which used in operational situations \cite{palmer_stochastic_2009,Berner2017}.}

{\mk Physically, getting to the limit \eqref{Hass1} allows for interpreting the  fast weather dynamics as inducing a diffusion process. As such, the understanding of the complex nonlinear interactions of e.g.~unstable modes with (possibly deep) stable ones \cite{chekroun2022transitions},  at the core of the chaotic nature of many geophysical flows  \cite{GhilChildress1987,dijkstra2005nonlinear,DG05} is replaced by the understanding of the interactions between noise and nonlinear effects \cite{CLW15_vol1,CLW15_vol2,chekroun2023transitions}.}

As appealing are such attributes, the  infinite timescale separation assumption underlying Eq.~\eqref{Hass1} is often challenged in climate applications. {\color{black}Wouters and Gottwald proposed {\mkr an extension of the MTV approach for the stochastic model reduction valid for} slow-fast systems with a moderate time scale separation \cite{Wouters_2019a,Wouters_2019b}.}
We discuss  below how to amend Hasselmann's program in such situations, either by seeking for natural extensions to Eq.~\eqref{Hass1} (Sec.~\ref{Sec_MZ}), or stochastic alternatives (Sec.~\ref{Sec_L80}). In the climate system, there is actually a multiplicity of spatio-temporal scales interacting across a wealth of processes (Fig.~\ref{Fig_Anna}) and the Hasselmann ansatz, whilst indeed inspiring, calls for its revision.

An important implication of Hasselmann's {\mk approach} is the provision of a probabilistic interpretation of climate dynamics, going beyond the study of individual trajectories. Indeed, the SDE \eqref{Hass1} can then be translated into a Fokker-Planck equation (FPE) \cite{pavliotisbook2014}  {\mk providing the probability distribution of the climate's states}, $\rho(\x,t)$, according to  
\be\label{eq:fpe 1}
\partial _t\rho= \mathcal{L}_0\rho=-\nabla \cdot  \lp \FF(\x)\rho \rp 
 +\frac{1}{2}\nabla \cdot \nabla \Big(\Sigmab \Sigmab^{T}(\x) \rho \Big).
\ee

In practice, $\rho(\x,t)$ is constructed using an ensemble of trajectories; see \cite{Maher2021} for a recent summary of the use of ensembles in climate modeling.  $\Sigmab \Sigmab^{T}$ is the non-negative definite noise covariance matrix.  
The unperturbed climate is then given by the stationary solution $\rho_0(\x)$ to the FPE defined by $\ellL_0 \rho_0 = -\nabla \cdot  \Big(\FF \rho_0\Big) +\frac{1}{2}\nabla \cdot \nabla \Big(\Sigmab \Sigmab^{T}\rho_0 \Big)=0$. {\color{black}Defining stationarity in a multiscale system is a nontrivial matter, as the stationary state can critically depend on the time-scale of interest. See {\bf Box 1} for a  discussion on this matter.} 
\begin{bclogo}{\small Box 1: Stationarity and Multiscale Behaviour}
  \footnotesize{The very notion of stationary reference climate is intimately related to considering autonomous dynamics as starting point of our analysis; see Eq.~\ref{fastslow}. But, instead, boundary conditions and external forcing relevant for the climate system do change on a variety of time scales \cite{vonderHeydt2021,Arnscheidt2022Feedbacks}; see a careful discussion of the very notion of statistical balance and stationary state for the climate system in a biogeochemical perspective in \cite{Arnscheidt2022Balance}. The  conundrum of defining stationarity in a multiscale system can be  pragmatically dealt with by taking advantage of Saltzman's viewpoint \cite{saltzman_dynamical}. One can extend Eq.~\eqref{fastslow} by including, instead of \textit{just} $(\x,\y)$, a larger set of qualitatively different variable $(\x_1,\ldots,\x_p)$  ordered according to their characteristic time scale, ranging from e.g.~\textit{very slow} to \textit{very fast} behaviour. Depending on the time scale of interest (e.g.~decadal vs multicentennial vs multimillenial  climate variability), one ends up choosing the relevant set of \textit{active} climatic variables of interest $\x_k$ (e.g.~atmosphere vs ocean vs ice caps, respectively), with {\mkr $\x_\ell$, $1\leq \ell <k$} determining the boundary conditions, and $\x_m$, $m>k$ contributing to defining the modified drift and noise law in Eq.~\eqref{Hass1}. Time-dependent forcing {\mkr terms} associated to geological of astrophysical processes can be considered as constant (or corresponding to stochastic forcing, in the case of e.g.~fast solar variability) when focusing on climatic processes of multimillennial or shorter time scales. Indeed, the very definition of climate system and of its evolution law depends on the time scale of interest; see \cite{GhilLucarini2020} for the related discussion of the so-called hierarchy of climate models. Hence, when referring to a reference stationary $\rho_0$, we are implicitly imposing a cutoff on the slow variability of the system.}
\end{bclogo}

When the noise term in Eq.~\eqref{Hass1} is sufficiently non-degenerate, i.e.~when, roughly speaking, the noise propagates out in the whole phase space
 through interactions with the nonlinear terms, the probability density $\rho_0$ is smooth; see e.g.~\cite[Appendix A.2]{Chekroun_al_RP2}. In contrast, when $\Sigmab=0$,  Eq.~\eqref{eq:fpe 1} becomes the Liouville equation and, in the case of dissipative chaotic systems, the  {\mk probability distribution $\rho_0$ is typically}  singular with respect to the Lebesgue volume in $\R^d$ \cite{eckmann_ruelle}.

Assuming that  $\int\mathrm{d}\x\rho_0(\x)=1$, the reference climatological mean for the observable $\Psi$ is $\langle \Psi \rangle_0= \int\mathrm{d}\x\rho_0(\x)\Psi(\x)$.  The function $\Psi$ could be in principle any quantity of climatic interest, corresponding to local properties or spatially averaged ones. It makes sense to associate possible $\Psi$'s with essential climate variables (ECVs), which are key physical, chemical, or biological variables  that critically contribute to the characterization of Earth's climate and are targeted for observations \cite{Bojinski2014}, or the quantities used for defining performance metrics for Earth System Models (ESMs) \cite{Eyring2020}. 

\subsubsection{Climate Response to Forcings}\label{responseHasselmann}
A second {\mk landmark} of the Hasselmann  program 
deals with the study of the response of climate models to perturbations. Aiming at studying how a climate model relaxes to steady state conditions or responds to perturbations, {\mk like} a sudden CO$_2$ increase, 
Hasselmann and collaborators heuristically proposed a methodology inspired by the dynamics of linear systems \cite{Maier-Reimer1987,Hasselmann1993}.  They showed that one can express the {\mk variation} $\delta \Psi(t)$ describing the departure of the model from steady state conditions (or convergence to it) by performing the convolution of a suitably defined causal Green's function {\color{black}$G_\Psi(t)$ ($G_\Psi(t)=0$ if $t<0$) with the time modulation of the acting perturbation $f(s)$:
\be\label{GHass}
\delta \Psi(t)=\int_{-\infty}^\infty \d s G_\Psi(t-s)f(s), 
\ee}
where $\Psi$ describes a climate variable of interest. This amounts to treating the problem of climate change 
using response theory. Leith pioneered this angle \cite{Leith1975} by proposing the use of the fluctuation-dissipation theorem (FDT) \cite{kubo1966} for expressing the Green's functions in terms of readily accessible correlations of climatic observables in the unperturbed state. Additionally, 
Hasselmann and collaborators expressed the Green's function $G_\Psi$ as a sum of exponential terms: 
\be\label{GHassExp}
G_\Psi(t)=\sum_k \alpha^k_\Psi \exp(\lambda^\Psi_k t),
\ee
where each of the $\lambda_k$ encodes an acting feedback.  
The Green's function-based method was shown to have good skill in performing climate change projections for individual model runs, after  filtering the natural variability \cite{Hasselmann1993} and 
for studying the carbon cycle in a climate model \cite{Maier-Reimer1987}. 
{\color{black}In Sec.~\ref{climateresponse} we  critically revise this approach by discussing how a formalism based on use of Green functions allows one to cast the problem of climate change as the response of the  system's probability distribution or of the statistics of the $\Psi$'s to (possibly time-dependent) perturbations applied to 
Eq.~\eqref{Hass1}  
\cite{GhilLucarini2020,Tel2020}. }

\subsubsection{Detection and Attribution of Climate Change}\label{DAHasselmann}
Using stochastic climate models one can make statements on climate {\color{black}variability and climate} change in terms of probability distributions, by computing averages and higher statistical moments. 
{\color{black}Separating climate variability and  climate change in the course of our single realization of the climate evolution relies on testing statistically departure from stationarity on observational time series. Performing causal attribution in the Pearl's sense of the climate change signal to specific forcings is a much harder task because it requires, as mentioned above, comparing information from observations and from climate model runs.  
In the  '90s Hasselmann and collaborators proposed the basic conceptual framework for performing attribution studies of climate change. 
\cite{Hasselmann1993b,Hegerl1996,Hasselmann1997}. This has played a major role in clarifying that we are presently experiencing a statistically relevant and physically attributable shift from previous climatic conditions \cite{Bindoff2013,IPCC2022}. Following \cite{Hannart2014}, the problem of attribution can be cast as follows: 
\begin{equation}
\label{eq:da3}
Y_k=\sum_{p=1}^F\tilde{X}_k^p\beta_p+\mathcal{R}_k \quad \tilde{X}_k^p={X}_k^p+\mathcal{Q}_k^{p}, k=1,\ldots,N.
\end{equation}
Here $Y_k$ is a vector describing the observed climate change. As an example, $Y_k$ could represent the 2011-2020 average temperature measured at $N$ different locations. The goal is to reconstruct it as a linear combination of $F$ regressors or fingerprints, i.e.~externally forced signals $\tilde{X}_k^p$, $p=1,\ldots,F$ associated with $F$ different forcing, plus a vector describing the natural variability of the system $\mathcal{Q}_k^{p}$\cite{Hasselmann1993b,Hegerl1996,Hasselmann1997,Allen1999,Allen2003}. The $p^{th}$ fingerprint $\tilde{X}_k^p$ is obtained as an ensemble average over possibly many simulations of the climate change signal obtained by perturbing the reference climate by applying exclusively the $p^{th}$ considered forcing. Yet, we can practically have access to the approximations ${X}_k^p$ for such fingerprints,} whereby the difference $\mathcal{Q}_k^{p}$ with respect to the true value is associated with our incomplete sampling of the model response and with model error. In many applications, information coming from different climate models is bundled together \cite{hegerl2011use}.

The goal is to perform an optimal inference of the  coefficients {\mk $\beta_p$} given the uncertainties described by  $\mathcal{R}_k$ and $\mathcal{Q}_k^p$, hence the expression \textit{optimal fingerprinting}. 
Usually the vectors column errors $\mathcal{R}_k$ and $\mathcal{Q}_k^p$, are modelled as independent, normally distributed stochastic vectors with zero mean and covariance matrices {\mk $\Cb$} and $\Omega_p$, respectively. 
The matrix $\Cb$ is constructed taking into account the correlations of the climatic variables in the unperturbed climate. 
A simpler  version of the theory above assumes $\Omega_p=0$ {\mk for all $p$} \cite{Allen1999}. In this case, via linear algebra one derives a relatively simple expression for the best estimates for the $\beta$'s {\mk along with} their  uncertainties. As a next step, if one assumes that the natural variability simulated by the climate models {\mk matches that of} the observations and one estimates  $\tilde{X}^p_k$ as the average of $L$ runs from a single climate model under the $p^{th}$ forcing, then, under the approximation above, one gets  $\Omega_p = \Cb/L$ \cite{Allen2003}.

One speaks of {\color{black}attribution 
 of climate change for the $p^{th}$ fingerprint if the confidence interval of $\beta_p$ does not intersect zero and includes the value one}. The procedure is truly successful if the confidence intervals for the $\beta_p$'s are not too spread out.  Optimal fingerprinting relies critically on the assumption of {\mk  linear dependence} of the climate response to the applied forcings. Additionally, the method, despite its great success, has recently been criticised as it has been suggested that uncertainties in the inference are sometimes underestimated \cite{Li2021}, or, more radically, on the basis that the statistical foundation of the procedure are not very solid \cite{McKitrick2022}; see also discussion in \cite{Chen2022}. {\color{black} In Sec.~\ref{optimalfingerprinting} below, we describe how the equations describing the Hasselmann's optimal fingerprinting method can benefit of new insights when reframed within the response theory of dynamical systems, and, in particular, can be derived seamlessly from linear response formalism.}

\subsection{This Review}

Our goal here is to give a critical appraisal of the Hasselmann programme based on some ideas and concepts in 
statistical mechanics and functional analysis {\color{black}that have gained traction}, for the most part, in the last two decades. 
We also propose a comprehensive framework for understanding the multiscale nature of climate variability and climate response to forcing and for fundamentally advancing our understanding of the ongoing climate crisis. This paper is structured as follows. 

Section \ref{modelreduction} {\color{black}discusses} 
rigorous theory-informed and effective data-driven model reduction strategies to highlight connections and integration between the top-down and bottom-up approaches, to relax the assumption of infinite time-scale separation discussed in Sec.~\ref{stochasticcm}. 
Reduced order modelling is inextricably associated with performing partial observations, i.e.~gaining only a partial, imperfect knowledge of the properties of a system.  As clarified by the Mori-Zwanzig (MZ) formalism \cite{mori_transport_1965,zwanzig_memory_1961}, the effective dynamics defining the evolution on the projected space of the variables of interest {\color{black}has} Markovian deterministic, stochastic, and non-Markovian components even if the dynamics of the whole system is purely deterministic; see e.g.~\cite{Chorin_al02,GKS04}. {\color{black}We review the different approaches for approximating the memory effects, and clarify the challenges posed by the presence of a modest} 
time-scale separation between the resolved and unresolved variables. 
{\color{black}We clarify physical situations in which the role of memory effects in the reduced model are secondary 
{\mkr whereas that of stochastic parameterization is key to recover the multiscale dynamics},
and we propose ways forward based also on the usage of neural parameterizations.} 

In Sec.~\ref{climateresponse} we show how response theory for nonequilibrium systems \cite{ruellegeneral1998,ruelle2009,Hairer2010,Baiesi2013,Sarracino2019,Gottwald2020,Santos2022} allows one to find explicit formulas and to devise experimental protocols aimed at   
performing climate change projections using climate models of different levels of complexity \cite{ragone2016,Lucarini2017,Aengenheyster2018,Lembo2020}. Hence, we shed light of some of the key aspects discussed in Sec.~\ref{responseHasselmann}. {\color{black}We clarify that the use of a formalism based on Green's functions does not requires assuming linearity of the model. Instead, one can linearize the statistical properties, as defined by ensembles averages, of the model around its reference steady state.} Taking advantage of the Koopman operator formalism \cite{Mezic2005,Budinisic2012,Kutz2016}, we show that it is indeed possible to write any Green's function as a weighted sum of exponentials, and we  carefully explain the meaning of the weights and of the decay rates. This angle  also facilitates understanding the basic properties of tipping points,  and  associating their presence to the divergence of the response operators \cite{Tantet2018,Chekroun_al_RP2,Santos2022}. 

In Sec.~\ref{optimalfingerprinting} we show how the linear response formalism developed in Sec.~\ref{climateresponse} provides the mathematical and physical backbone behind the optimal fingerprinting method for detection and attribution of the climate change signal presented in Sec.~\ref{DAHasselmann}. Additionally, our angle allows one to better appreciate the approximations taken in the practice of detection and attribution studies (especially regarding the definition of the error terms), clarify some fundamental issues associated with the use of mixtures of different climate models in {\color{black}defining} the fingerprints, and elucidate why the fitting strategy might fail in the proximity of tipping points. This allows to convincingly prove the strong link between the three  main axes of Hasselmann's research programme. 

In Sec.~\ref{conclusions} we present our conclusions and perspectives for future work.

\section{Theory-Guided and Data-driven Model Reduction}\label{modelreduction}

{\mkr We present in this section the quest for stochastic model reduction, in a more general setting than for the slow-fast systems discussed in Sec.~\ref{stochasticcm}, as framed here within the closure problem from the point of view of the Mori-Zwanzig (MZ) expansion. We review in Secns.~\ref{Sec_MZ} and \ref{Sec_variational} below the mathematics behind this expansion, and review the different approaches adopted in the literature to approximate the elements of this expansion in particular regarding the memory and stochastic terms. We argue   however that, unlike common belief, there are physically-relevant regimes for which the three groups of terms do not share the same predominance, even in presence of no timescale separation.  Section \ref{Sec_L80}  gives an important example tied to Primitive Equations, the fundamental equations of atmosphere-ocean dynamics, in regimes for which the absence of timescale separation is manifested in each of the model's variables by bursts of fast dynamics popping out irregularly in time on top of a slow trend motion.  There, we review that the proper knowledge of the manifold capturing this slow trend motion (tied to Rossby waves) enables us to figure out that the dynamics off this manifold is tied to inertia-gravity waves and that the latter can be efficiently parameterized by means of coupled stochastic oscillators, without the need of memory terms.} {\mkr Another example is reviewed in Sec.~\ref{Sec_ML_turb} for the closure problem of  forced two-dimensional turbulence. There, we review that the recent advances on machine learned parameterizations allow for having access of accurate coarse-grained closures without the need of memory and even noise terms. Of course, the latter example is subject to the choice of the cutoff scale, as lowering the latter will inevitably be prone at some point to the emergence of such terms, referring thus to the material reviewed
in Secns.~\ref{Sec_MZ} and \ref{Sec_variational}. We finally illustrate in Sec.~\ref{Sec_too_manyfreq}, on a data-driven modeling problem of coarse-grained oceanic turbulence, the importance of the choice of latent variables for simplifying the equation discovery problem {\it a la MZ} from time-sequential data that have large spatial dimensions.}

\subsection{Mori-Zwanzig decomposition from perturbation theory of the Koopman semigroup}\label{Sec_MZ}
We consider a dynamical system {\mkr of the form
\be\label{EqF}
\dot{\U}=\Fb(\U), \;\; \U\in \mathbb{R}^N.
\ee
The state vector $\U$ can be assumed here to be high-dimensional, describing a the state of a group---or subgroup---of climate variables (atmospheric variables, streamfunctions, etc.). Eq.~\eqref{EqF} can be thought as resulting from discretization of a system of partial differential equations (PDEs) describing the motion of geophysical fluids or the evolution of other climate variables \cite{GhilChildress1987}. 
As such, \eqref{EqF} does not necessarily belongs to the class of slow-fast systems such as \eqref{fastslow}, although the latter can be recast into the general abstract setting presented here. We denote by $S(t;\U)$  the solution to Eq.~\eqref{EqF} emanating from $\U$ in $\mathbb{R}^N$ at some initial time.}

{\mkr We assume that long-term statistics such as power spectral densities or correlations are well defined from Eq.~\eqref{EqF}, namely that Eq.~\eqref{EqF} possesses an invariant probability distribution $\mu$, also known as stationary statistical equilibrium, that satisfies:
\be\label{Eq_phys_rev}
\underset{T\rightarrow \infty}\lim \frac{1}{T} \int_0^{T}  {\mkr \varphi(S(t;\u)) \d t =\int  \varphi( \U ) \d \mu ( \U)}, 
\ee
for almost  every  $\U$ (in the Lebesgue sense)}  that lies in the basin of attraction of $\mu$, for any sufficiently smooth observable $\varphi$. In general $\varphi$ is a field quantity which represents perturbations of, e.g., density, pressure, electrostatic potential, etc. Note that when a  global attractor exists, a statistical equilibrium $\mu$ satisfying \eqref{Eq_phys_rev} is  supported by the global attractor (see e.g.~\cite{CGH12}) {\mkr and gives the asymptotic ``mother'' distribution of the dynamics over this attractor from which any probability density functions (PDFs)  are derived (through e.g.~marginals)}. {\color{black}.We remark that such statistical equilibrium should not be confused with the classical notion of thermodynamical equilibrium in statistical mechanics.}

Given an observable $\varphi: \mathbb{R}^N \rightarrow \mathbb{R}$, recall that the evolution of this observable along the flow associated with Eq.~\eqref{EqF} is given by the Koopman semigroup, $K_t$, defined as \cite{Budinisic2012} 
\be\label{Eq_Koopman_op}
{\mkr K_t \varphi(\U) =\varphi(S(t;\U)),}
\ee
that satisfies the Liouville equation $ \partial_t K_t \varphi=\mathcal{L} K_t \varphi$
 with {\mkr the operator $ \mathcal{L} $}
 \be\label{Def_L}
 \mathcal{L} \varphi = \Big(\sum_{j=1}^N {\mkr F^j (\U)} \partial_j\Big) \varphi,
 \ee
 denoting the Lie derivative along the vector field $\bm{F}$, {\mkr while the $F^j$ denote the components of the latter}.

We are now given a decomposition {\mkr $\U=(\u,\vb)$ in which $\u$ denotes a vector of $p$ relevant variables, those resolved typically, while $\vb$ denotes the vector collecting the} $q=N-p$ neglected ones.
We are interested in describing the evolution of any sufficiently smooth observable $\psi$ of the variable $\u$ {\mkr only}, without having to resolve the $\vb$-equation in Eq.~\eqref{EqF}. In climate dynamics, this {\mkr problem} motivated for instance by predicting/simulating a scalar field of interest (e.g.~temperature, pression) over a coarse grid without having to resolve the subgrid processes (closure problem). {\mkr Within this context, one may think of $\u$ as a collection of coarse-scale variables, and $\vb$ as a collection of small-scale ones. In the case Eq.~\eqref{EqF} is an abstract formulation of  a slow-fast system, the variable $\u$ (resp.~$\vb$) will be referred as slow (resp.~fast) in what follows}, {\color{black}thus corresponding to the $\x$ and $\y$ variables in Eq.~\eqref{fastslow}, respectively.}

In operator form,  
{\mkr given an observable $\psi$ of the coarse-scale variable $\u$ only, the problem consists of finding a parameterization in following transport equation of  the coefficients depending on unresolved variable $\vb$, 
\bea\label{Eq_Liouville}
\partial_t \psi =\Big(\sum_{j=1}^p F^j \big(\u(t),\vb(t)\big) \partial_j\Big) \psi,
\eea
in order to be able to describe the time-evolution of $\psi(\u(t))$ without having to resolve the evolution of the $\vb$-variable.}
This PDE describing how {\mkr any observable $\psi$ of the reduced state space is advected by the flow of Eq.~\eqref{EqF} is obtained by observing that 
 $\partial_t \psi(\u(t))=\nabla_{\u} \psi \cdot \partial_t \u $}. 
{\mkr Thus, the key issue we aim at solving is an effective parameterization of the interactions between the coarse-scale and subgrid variables $\vb$ (via the terms $F^j$) for an accurate description of the advection of $\psi$ in terms of the resolved variable $\u$, only}.   

To address this closure problem, {\mkr associated with the cutoff scale consisting of retaining the $\u$-variable,} we introduce the conditional expectation operator {\mkr acting on observables of the full state vector $\U=(\u,\vb)$}
\be\label{Eq_averaging_ope}
{\mkr  \big[ P \varphi \big] (\u) = \int \varphi(\u,\vb) \d \mu_{\u} (\vb)=\overline{\varphi}},
\ee
in which $\mu_{\u}$ denotes the disintegration of the invariant probability measure $\mu$ \cite[Theorem 5.3.1]{ambrosio2008gradient}; roughly speaking it gives the distribution of the $\vb$-variable on the attractor when the {\mkr coarse-scale variable is frozen to $\u$.}
The operator $P$ corresponds thus to an averaging with respect to the neglected variable $\vb$ as conditioned on $\u$. Note that  $\mu_{\u}\neq \mu_{\u'}$ for $\u\neq \u'$ for complex systems causing the variable $\vb$ not to be   identically and independently distributed (i.e.~not i.i.d.). 
 
 Now, by rewriting the transport equation \eqref{Eq_Liouville} as $ \partial_t (K_t \psi ) =K_t \M \psi$ with $\M=\sum_{j=1}^p F^j  \partial_j$, we have that  
 \be\label{Eq_basic_pert}
 \partial_t (K_t \psi) =K_t P \M \psi+K_t (\M \psi -P\M \psi).
 \ee 
 Here, $ P\M =\sum_{j=1}^p \overline{F^j}(\u) \partial_j$
 denotes the averaged advection operator with respect to the neglected, unresolved variable $\vb$.

By performing the change of variable $s \leftarrow t-s$ in the integral term of Eq.~\eqref{Eq_f0} in {\bf Box 2}, we arrive finally at the following equivalent formulation of Eq.~\eqref{Eq_f0}
{\mkr 
\bea\label{Eq_GLE}
 \partial_t \psi&=P \M \psi (\u(t)) \\
& \qquad+\hspace{-.2ex}\int_0^t  \Gamma(t-s)  \psi (\u(s)) \d s +\eta(t)\psi,
\eea
}
which gives the desired closure of Eq.~\eqref{Eq_Liouville}.

Eq.~\eqref{Eq_GLE} is {\mkr called} the  {\it Generalized Langevin Equation (GLE)}  \cite{pavliotisbook2014} or the {\it Mori-Zwanzig (MZ) decomposition}. {\mkr Thus, the} effective dynamics defining the evolution of any observable 
of the reduced state space can be achieved by determining the 
Markovian, stochastic, and non-Markovian components appearing in Eq.~\eqref{Eq_GLE} making the GLE, the fundamental equation to determine in the Mori-Zwanzig approach to closure 
\citep{zwanzig_memory_1961,mori_transport_1965,Chorin_al02,GKS04,Chorin_Hald-book}. We  {\mkr clarify below  a subtle point for applications, namely that not all the terms are necessarily} needed  in this triptych decomposition {\mkr to derive accurate closures for multiscale dynamics even when no timescale separation is apparent. {\color{black}In the special case of  one-way coupling between the subgrid and resolved variables the non-Markovian component mentioned above disappears \cite{Vissio2018c}.} {\mkd Section \ref{Sec_L80} below reports on a fully-coupled model from atmospheric dynamics, where the non-Markovian terms are negligible but the stochastic ones are determining to recover the  multiscale nature of the dynamics from closure}.

{\mkdd  MZ decompositions have attracted a lot of attention in the last two decades as a promising description for reduced modeling of coarse-grained variables \cite{GKS04,Chorin_Hald-book,hijon2010mori}, in many areas such as molecular dynamics \cite{izvekov2006modeling,chen2014computation,ma2016derivation}, climate dynamics \cite{chekroun2011predicting,Majda_Harlim2012,wouters2013multi,ghil2015collection,MSM2015,chekroun2017data,Boers_al17,falkena2019derivation}, or fluid problems \cite{parish2017dynamic,parish2017non,wang2020recurrent} to name a few. }

  \begin{bclogo}{\small Box 2: Derivation of the GLE}
  \footnotesize
 In Eq.~\eqref{Eq_basic_pert}, the operator,
$$\M  -P\M=\sum_{j=1}^p \big(F^j (\u,\vb)-\overline{F^j}(\u)\big) \partial_j,$$ accounts for the fluctuations with respect to the conditional average and $K_t (\M \psi -P\M \psi)$ informs thus on how these fluctuations are transported by the flow of Eq.~\eqref{EqF} {\mkr for any observable $\psi$ of the reduced state space}.    
Hence, the operator  $\bm{\delta}f = (Id- P)\M  f= Q \M f$
encodes the fluctuations terms.  It defines the fluctuation semigroup $e^{t \bm{\delta}}$ {\mkr that constitutes a key element in the closure of Eq.~\eqref{Eq_Liouville}. This closure is indeed obtained} by application of the perturbation theory  of semigroups  in the Miyadera-Voigt  variation-of-constants formulation \cite{miyadera1966perturbation,voigt1977perturbation,engel2000}, {\mkr this fluctuation semigroup and the Koopman operator given by \eqref{Eq_Koopman_op}. The Miyadera-Voigt formula \eqref{Eq_MZ_decomp} below, is also known as the Dyson's formula in the MZ literature \cite{Chorin_al02}.}

The  Miyadera-Voigt perturbation {\mkr theorem \cite[Sec.~3.c]{engel2000}} gives then that
\be\label{Eq_MZ_decomp}
K_t  f =e^{t\bm{\delta}} f +\int_{0}^t K_{t-s} P\M e^{s \bm{\delta}} f \d s,
\ee 
for any observable $f$  for which $\bm{\delta} f $ is well defined. Note that $P e^{t \bm{\delta}} f=0$ if $P f =0$, i.e.~the {\it orthogonal complement}  of $P$,  is invariant under $e^{t\bm{\delta}}$. Note that the fluctuation semigroup $e^{t \bm{\delta}}$ gives the solution of the orthogonal dynamics equation
\be\label{Eq_ortho}
\partial_t e^{t \bm{\delta}} f= Q \M e^{t \bm{\delta}} f.
\ee
We refer to \cite{givon2005existence} for the study of existence of solutions to Eq.~\eqref{Eq_ortho}. 
 
Now let us take $f=\M \psi-P\M \psi=Q \M \psi$ in \eqref{Eq_MZ_decomp} and observe that $P f=0$. Then Eq.~\eqref{Eq_basic_pert} becomes:
\bea\label{Eq_f0}
\partial_t K_t \psi &=K_t P \M \psi+e^{t \bm{\delta}} f +\int_{0}^t K_{t-s} P\M e^{s \bm{\delta}}  f  \d s\\
&=K_tP \M \psi\hspace{-.2ex} +\hspace{-.2ex}\int_0^t \hspace{-.2ex} K_{t-s} \Gamma(s) \psi  \d s +\eta(t)\psi,
\eea
with  $\eta(t)\psi=e^{t \bm{\delta}} Q\M \psi$
denoting the orthogonal element of the MZ decomposition (since   $P Q \M \psi=0$ implying that $\eta(t)\psi$ lies in ker($P$)), while 
$\Gamma(s)$ defines the  operator $\Gamma(s)=P\M \eta(s)$
\end{bclogo}

In Eq.~\eqref{Eq_GLE}, the  {\mkr kernel operator}  $\Gamma(t-s)$ is typically a time-lagged damping kernel and $\eta(t)$ is interpreted as an effective random forcing uncorrelated with the time-evolution of the {\mkr resolved variable, $\u(t)$,} but can be strongly correlated in time; see Sec.~\ref{Sec_L80} below.  In the slow-fast system metaphor, the Markovian term $P \M$ provides the slow component of the dynamics,  $\eta(t)$ is void of slow oscillations, while $\Gamma$ is supposed to account for the disparate interactions between the timescales.  

As elegant it may be,  the MZ decomposition is a technically challenging solution to the closure problem of disparate scale interactions and various assumptions about the memory kernel $\Gamma$ are typically made to propose approximations to the GLE. 

The memory kernel  $\Gamma$ and the "noise" operator $\eta(t)$ involve the implicit knowledge of the fluctuation semigroup  $e^{s \bm{\delta}}$, accounting for the effects of the neglected variables on the fluctuations with respect to the average  motion. This operator is difficult to resolve as it boils down of solving the orthogonal dynamics equation Eq.~\eqref{Eq_ortho}  \cite{stinis_Higher-order}. 

The noise and memory terms can be extremely complicated to calculate, especially in cases with weak or no obvious timescale separation between the resolved and unresolved variables. The approximation of these terms constitutes thus the main theme of most research on the MZ decomposition.

Many techniques have been proposed to address this problem in practice and can be grouped in two categories: (i) data-driven methods, and  (ii) methods based on  analytical insights tied to the very derivation of the MZ decomposition. Data-driven methods aim at recovering the MZ memory integral and fluctuation terms based on data, by exploiting sample trajectories of the full system.
 Data-driven methods can yield accurate results, but they often require a large number of sample trajectories to faithfully capture memory effects \cite{li2015incorporation,lei2016data,li2017computing,brennan2018data}.
Typical examples include the NARMAX (nonlinear auto-regression moving average with exogenous input) technique developed by \cite{chorin2015discrete,lu2017data,Lin.Lu.2021}, the rational function approximation  proposed in \cite{lei2016data}, the conditional expectation techniques of \cite{brennan2018data}, methods based on Markovian approximations  by means of {\mkdd surrogate hidden variables \cite{Majda_Harlim2012,MSM2015,lei2016data,harlim2021machine,qi2023data}, or kernel-based linear estimators in delay-coordinate \cite{gilani2021kernel}.} 

Methods based on analytical considerations aim at approximating the MZ memory integral and fluctuation terms based on the original {\mkr model's} equations, without using any simulation data. The first effective method developed within this class can be traced back to the continued fraction expansion of Mori \cite{mori1965continued}, which can be conveniently formulated in terms of recurrence relations \cite{lee1982solutions,florencio1985exact}; see also \cite{kupferman2004fractional}. Other {\mkr theoretically-guided} methods to compute the memory and fluctuations terms in the MZ decomposition include optimal prediction methods  \cite{Chorin_al_2002,chorin2007problem,Stinis06}, mode coupling techniques \cite{gotze1999recent,reichman2005mode}, methods based on approximations of the orthogonal equations \cite{darve2009computing}, matrix function methods \cite{chen2014computation}, series expansion methods \cite{stinis2015renormalized,parish2017non,parish2017dynamic, zhu2018estimation,zhu2018faber}, perturbation methods \cite{venturi2014convolutionless}, and methods based on Ruelle's response theory  \cite{wouters2012,wouters2013multi}. These analytically grounded methods can lead to the effective calculations of the non-Markovian effects in various applications such as e.g.~in coarse-grained particle simulations \cite{yoshimoto2013bottom,hijon2010mori} or some fluid problems \cite{parish2017dynamic,parish2017non}, including intermediate complexity climate models \cite{Demaeyer2018}. However, these calculations are often quite involved and they do not generalize well to systems with no scale separation \cite{GKS04}; see, instead, an example of scale adaptivity in \cite{Vissio2018a}. 

In fact to better appreciate the difficulty posed by the lack of timescale separation it is useful to recall that for instance long-range memory approximation consisting of keeping the zeroth order term in a Taylor expansion of the memory {\mkr operator} in Eq.~\eqref{Eq_GLE} allows for simplifying significantly the memory term calculation, but at the price of  restrictive conditions. Indeed, such a long-range approximation shows relevance if the  unresolved modes exhibit sufficiently slow decay of correlations ($t$-model \cite{hald2007optimal,chorin2007problem}), essentially by assuming information about initial value to be sufficient to make predictions. Assuming the unresolved modes to have fast decay of correlations, one is left with short-range memory approximation schemes. 
As was shown in  \cite{chorin2007problem}, the two cases, of extreme or very weak non-locality in time, are the two sides of the same coin. Most of the challenging cases for closure lies thus in the intermediate
cases \cite{Stinis06}, for which there is no neat separation of timescales such as populating climate science \cite{GhilLucarini2020}.

Keeping higher-order terms in the Taylor expansion of the memory {\mkr operator} is a natural way to handle cases of weak timescale separation. It is illuminating in many ways, including to design data-driven methods, as explained below.
This   higher-order  approximation approach of the memory {\mkr operator} has been retained by Stinis in \cite{stinis_Higher-order} and further developed and analyzed by Zhu et al.~in \cite{zhu2018estimation}. The approach consists of breaking down the memory approximation problem into a  hierarchy of auxiliary Markovian equations.

Denoting by $\m_0(t)$ the integral term in Eq.~\eqref{Eq_GLE}, such a Markovian approximation is accomplished by observing  that $\m_0(t)=\int_0^t K_s P \M e^{(t-s) Q\M} Q \M$ satisfies the following  infinite-dimensional system of PDEs \cite{stinis_Higher-order,zhu2018estimation}
\be\label{Eq_Hmodel}
\frac{\d \m_{n-1}}{\d t}= K_t P\M (Q\M)^nv+\m_{n}(t),\quad n=1,\ldots.
\ee
Integrating Eq.~\eqref{Eq_Hmodel} backward, i.e., from the ``last'' equation to the first one, to obtain  a Dyson series representation of $\m_0(t)$ involving repeated integrals \cite{zhu2018estimation}.  
In practice, one performs a truncation of Eq.~\eqref{Eq_Hmodel}, which consists of keeping the first $n$ equations, while closing the last equation by using an ansatz in place of  $\m_{n}(t)$, such as $\m_{n}(t)=0$   \cite{stinis_Higher-order} or  $m_n(t)$ given by Chorin's $t$-model; see \cite{zhu2018estimation} for other choices.  Depending on the the order of truncation retained and the corresponding choice of the ansatz for $\m_n$,  error estimates with respect  to the genuine memory integral in Eq.~\eqref{Eq_GLE} are available  \cite{zhu2018estimation}. 
The implementation of such Markovian schemes is however not trivial to conduct as it requires computing $(Q\M)^n$ to a high-order in $n$,  a delicate operation to accomplish especially when the original system is large, recalling that $Q\M$ is the generator of the orthogonal equation \eqref{Eq_ortho}.

Nevertheless, the layered structure of  Eq.~\eqref{Eq_Hmodel} and related error estimates provide a strong basis for the design of data-driven methods based on Markovianization ideas to approximate the memory integral term. We mention that such ideas are commonly used for the mathematical analysis of physical models involving integro-differential equations; see e.g.~\cite{Chek_al11_memo,chekroun_glatt-holtz} and references therein.  

The 
 data-driven approach proposed initially by Kravtsov et al.~in \cite{kravtsov2005multilevel} is intimately related to such Markovianization ideas for the MZ decomposition, as pointed out  in \cite{MSM2015}. The class of data-driven models of  \cite{kravtsov2005multilevel} involves also mulitlayered SDEs of a structure very similar to that of Eq.~\eqref{Eq_Hmodel} that has been generalized in  \cite{MSM2015} to handle the approximation of more complex memory kernels, from a data-driven perspective. The usage of such mulitlayered SDEs to provide approximation of the GLE is further discussed in Sec.~\ref{Sec_variational} below.

Efforts to approximate the memory and noise terms should not, however, make us lose sight of another key problem, namely the problem of approximating the conditional expectation, namely the Markovian terms in Eq.~\eqref{Eq_GLE}. This is where recent hybrid approaches exploiting the {\mkr original model's equations} and simulated data have shown relevance {\color{black} and have indicated that a blind application of data-driven methods can lead to uncertain outcomes}  {\mkd or incorrect interpretations even when the latter are successful; cf.~\cite[Sec.~6]{CLM19_closure} and \cite[Sec.~4.1]{ma2019model}}.}  

In that respect, the data-informed and theory-guided variational approach introduced in \cite{CLM19_closure}  allows indeed for computing approximations  of the conditional expectation term, $P\M$, by relying on the concept of the optimal parameterizing manifold (OPM) \cite[Theorem 5]{CLM19_closure}.  The OPM is the manifold that averages out optimally the neglected variables as conditioned on the resolved ones \cite[Theorem 4]{CLM19_closure}. 
The approach {\mkr introduced in \cite{CLM19_closure}  to determine OPMs in practice  consists to first derive}  analytic parametric  formulas that match rigorous leading approximations of unstable/center manifolds or slow manifolds near e.g.~the onset of instability, and {\mkr then to perform a data-informed minimization of a least-square parameterization defect in order to recalibrate} the manifold formulas' parameters to handle regimes further away from that instability onset  \cite[Sec.~4]{CLM19_closure}.    
{\mkr There, the optimization stage allows for alleviating the small denominator problems \cite{Arnold88} rooted in small spectral gaps \cite[Remark III.1]{CLM23_OPM_transitions}, and derive thereby accurate parameterizations in regimes where constraining spectral gap or timescale separation conditions are responsible for the well-known failure of standard invariant/inertial or slow manifolds \cite{debussche1991inertial,temam2011slow,zelik2014inertial}; see \cite[Sec.~6]{CLM19_closure} and \cite[Sec.~V]{CLM23_OPM_transitions} for examples.}
{\mkd In more physical terms, this problem is also tied to deficient or excessive parameterization of the small-scale energy but dynamically important variables, leading to an incorrect reproduction of the backscatter transfer of energy to the large scales \cite{kraichnan1976eddy,leith1990stochastic,debussche1995nonlinear}, and to inverse cascade errors \cite{debussche1995nonlinear,dubois1998incremental,dubois1998dynamic}.}

 For multiscale dynamics,  failure in resolving accurately the conditional expectation results typically into a residual that contains too many spurious frequencies 
to be efficiently resolved by data-driven methods based e.g.~on the aforementioned multilayered SDEs, the latter exploiting either polynomial libraries of functions or other specified interaction laws \cite{MSM2015} between the resolved and unresolved variables.

Lately, much efforts relying on machine learning (ML) techniques have been devoted for the learning of memory terms in MZ-decompositions  \citep{fu2020learning,wang2020recurrent,gupta2021neural,harlim2021machine,qi2023data}. These go beyond prior efforts involving 
polynomial libraries of specific interaction laws between the slow and fast variables \citep{wouters2013multi,kravtsov2005multilevel,MSM2015}. 
{\mkdd With the ML power one may be tempted though to use complex neural architectures to learn the MZ terms, but this should not be done at the price of  physical interpretations and understanding. Careful studies in that regard include \cite{harlim2021machine,qi2023data}. The examples discussed below provide other elements for caution.}

\subsection{Variational approach to closure}\label{Sec_variational}

In the context of subgrid parameterizations, {\it nonlocality} in time in the GLE \eqref{Eq_GLE} means that the subgrid variables exert reactive as well as resistive forces on the resolved variables, and as noted in \cite{kraichnan1987eddy} this may play an important role in reproducing finite-amplitude instabilities and other properties of these variables. 
{\mkr In absence of time-scale separation, the subgrid variables exert fluctuating driving forces on the resolved variables which are conceptually distinct from eddy viscosity (or even negative eddy viscosity) \cite{rose1977eddy}.}


We assume thus that $F$ in Eq.~\eqref{EqF} proceeds from a forced fluid model, i.e.~that {\mkr $F(\U)=L \U + B(\U,\U)+\bm{f}$} with $B$ denoting a bilinear operator, $L$ a linear {\mkr dissipative} operator, $\bm{f}$ a {\mkr time-independent} force {\mkr (to simplify)}, and {\mkr $\U=(\u,\vb)$ as in Sec.~\ref{Sec_MZ} above}. We are interested in finding an accurate closure in the slow/coarse-scale $\u$-variable. To achieve this goal, the parameterization of the $\u$-$\vb$ and $\vb$-$\vb$ interaction-terms in the original $\u$-equation,  i.e.~the terms accounting for the disparate-scale and fast-scale interactions, is the key issue.  Denoting by $\boldsymbol{\tau}_{int}$ the grouping of these interaction terms, a convenient way to address this problem is by seeking for {\mkr the nonlinear vector field $\boldsymbol{\tau}_{opt}$ of the $\u$-variable} that solves the minimization problem 
\be\label{Min_tau_int}
\underset{\boldsymbol{\tau}}\min  \int_0^{T} \bigg\| \boldsymbol{\tau}(\u(t))- \boldsymbol{\tau}_{int}(\u(t),\vb(t)) \bigg\|^2 \mathrm{d} t, \; T\gg 1.
\ee
The optimal parameterization, $\boldsymbol{\tau}_{opt}$, relates naturally to the conditional {\mkr expectation \eqref{Eq_averaging_ope} $\bm{F}$ since $\boldsymbol{\tau}_{opt} (\u) =  \overline{\bm{F}}$} minus the linear and $\u$-$\u$ interaction  terms that project onto the coarse-scale variables.

The aforementioned OPM, $\Phi_{opt}$, providing the best approximation in a least-square sense of $\vb$ as a mapping of $\u$, satisfies then that 
$\boldsymbol{\tau}_{int}(\u,\Phi_{opt}(\u)) \approx  \boldsymbol{\tau}_{opt}(\u)$,
with a small residual error when the $\vb$-$\vb$ interaction terms are negligible after averaging in the original $\u$-equation (such as \cite[Assump.~A4]{majda2001mathematical}); 
see \cite[Theorem 5]{CLM19_closure}. At this stage, knowing $  \boldsymbol{\tau}_{opt}$ or $\Phi_{opt}$  allows us thus to approximate the  average motion of {\mkr $\u(t)$} when averaging is performed  over the {\mkr unresolved variable $\vb(t)$}.
  
If one wants to recover beyond averaging,  the effects of the {\mkr fluctuations carried by $\vb(t)$} onto  the dynamics of {\mkr $\u(t)$}, then the MZ formalism recalled in Sec.~\ref{Sec_MZ} invites us  to revise the minimization problem \eqref{Min_tau_int} {\mkr as follows
\begin{widetext}
\be\label{Eq_min_memory}
\underset{\boldsymbol{\tau},\Gamma } \min  \int_0^T\bigg\| \boldsymbol{\tau}(\u(t))+ \int_0^t \Gamma (t-s) \u(s)\d s - \boldsymbol{\tau}_{int}(\u(t),\vb(t)) \bigg\|^2 \mathrm{d} t, \; T\gg 1.
\ee
\end{widetext}
}

Solving this second minimization problem consists thus of decomposing 
the nonlinear interaction term  to account for a memory function and a fluctuating force, {\mkr namely
\bea\label{Min_tau_int2}
\boldsymbol{\tau}_{int}(\u(t),\vb(t)) \approx &\boldsymbol{\tau}(\u(t))\\
&\;+ \int_0^t \Gamma (t-s) \u(s)\d s +\bm{r}(t),
\eea
with $\bm{r}(t)$ denoting the residual obtained after minimization of \eqref{Eq_min_memory}. This minimization} can be addressed by means of recurrent neural networks {\mkr allowing for functional dependence of the ``past'' such as} long short-term memory {\mkr (LSTM) networks} \cite{fu2020learning,wang2020recurrent,gupta2021neural}. {\mkr Although allowing for learning possible complex functional dependences in $\bm{\tau}$ and of the operator $\Gamma$, the resulting learned} elements {\mkr via neural networks} suffer from interpretability

Another approach consists of pursuing the minimization of \eqref{Eq_min_memory} via Markovianization which consists of breaking down the memory terms and noise terms by means of SDEs with a multilayer structure (similar to Eq.~\eqref{Eq_Hmodel}) whose coefficients are learned successively via recursive regressions using surrogate, stochastic, variables that account for the residual errors produced by the successive regressions until a white noise limit is reached  \cite{Majda_Harlim2012,MSM2015}. 
This data-driven approach \cite{kravtsov2005multilevel} has led to striking results in many fields of applications such as for the modeling of El-Ni\~no-Southern Oscillation (ENSO) \cite{kkg05_enso,ckg11,chen2016diversity}, extratropical atmospheric dynamics \cite{kkg06}, paleoclimate \cite{Boers_al17}, or the {\mkr Madden-Julian Oscillation \cite{kcg_13MJO,chen2014predicting,chen2015predicting}} to name a few. 

These regression-based multilayered SDEs to approximate the MZ decomposition Eq.~\eqref{Eq_GLE} benefit furthermore from useful theoretical insights. Indeed, intimate connections with the multilayered SDEs  derived in \cite{wouters2012,wouters2013multi} based on Ruelle's response theory \cite{santos2021reduced}, were shown to hold  for a subclass of   multilayered SDEs considered in \cite{Majda_Harlim2012,MSM2015}; see \cite{santos2021reduced}.  These connections allow in particular for clarifying circumstances of success for multilayered SDEs with linear coupling terms between the layers corresponding to approximating the memory {\mkr operator} $\Gamma$ in Eq.~\eqref{Eq_GLE} by repeated convolutions of exponentially decaying kernels  \cite{MSM2015}. The multilayered SDEs of this form were shown to be particularly relevant for weakly coupled slow-fast systems  and the corresponding   memory and noise terms were shown to relate naturally to the Koopman eigen-elements of the ``unperturbed weather'' Koopman semigroup $K_t^w$ whose generator is {\mkr $\sum  G^j(\vb) \partial_j$  when $\bm{g}(\u,\vb)=\bm{G}(\vb)+\epsilon C(\u,\vb)$} in Eq.~\eqref{fastslow} with $\epsilon$ small; see \citep[Theorem 2.1]{santos2021reduced}. The approximation of the MZ decomposition Eq.~\eqref{Eq_GLE} via Markovianization sheds thus new lights onto Koopman modes \cite{Budinisic2012} and related dynamic mode decomposition (DMD), widely praised in fluid dynamics over the last decade \citep{Rowley2009, Schmid2010,Kutz2016}, and as such with the Principal Oscillation Pattern modal proposed earlier in atmospheric sciences by Hasselmann \citep{Hasselmann.POPs.1988, Tu.ea.2014}.

However as mentioned earlier,  it is not always required, depending on the problem, to determine the memory and/or noise terms, and we should thus always look first for the virtue of solving the minimization problem \eqref{Min_tau_int} in the first place instead of solving the more 
challenging {\mkr minimization problem  \eqref{Eq_min_memory}}, which may involve memory or  noise terms of negligible importance for closure {\mkr depending e.g.~on the cutoff scale retained for a given dynamical regime; see Sec.~\ref{Sec_ML_turb} below}.

 An emblematic example is found {\mkr also in} the context of the Primitive Equations of the atmosphere. It is known  that at low Rossby number, the conditional expectation coinciding with the Balance Equation  is amply sufficient for an accurate closure \cite{chekroun2017emergence}.  However, once a critical Rossby number is crossed, the Balance Equation needs to be seriously amended to capture the complex interactions between the Rossby waves and  inertia gravity waves; the latter becoming non-negligible at large Rossby number, see below. 
\begin{figure*}
\includegraphics[width=0.95\textwidth, height=0.35\textwidth]{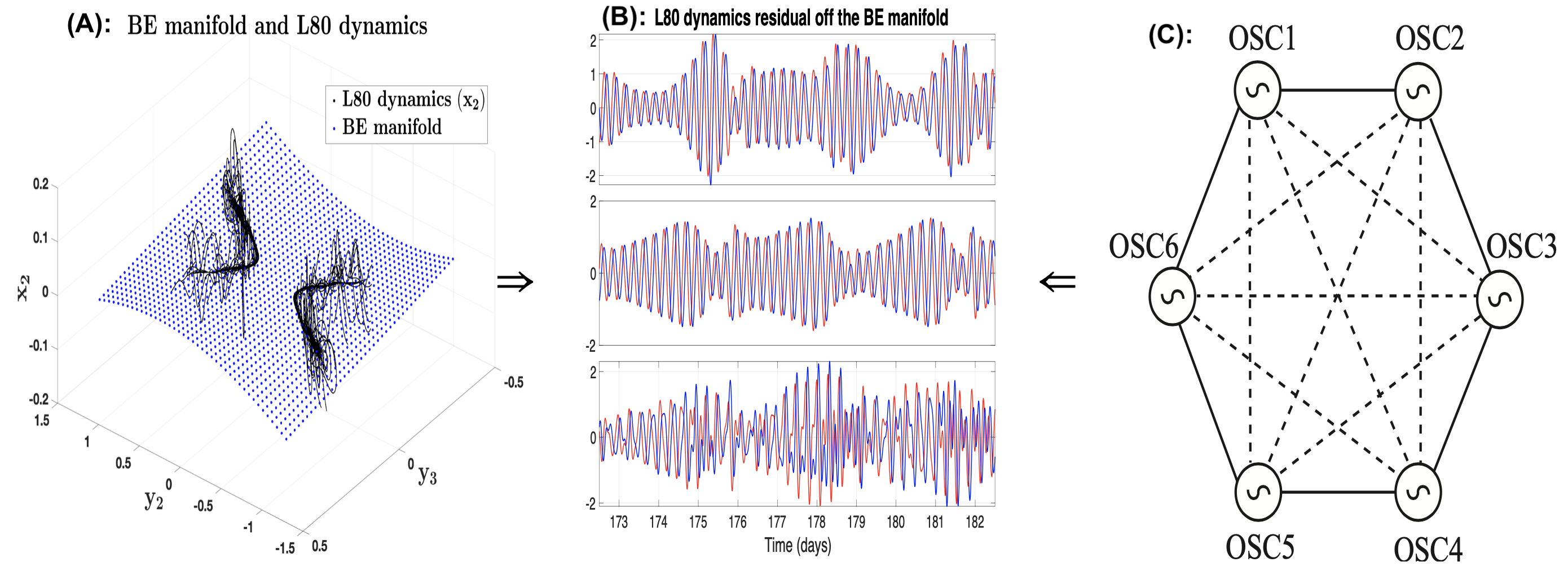}
\caption{\small {\bf An example of no need of memory but noise term in the MZ-decomposition Eq.~\eqref{Eq_GLE}.} Example from the atmospheric Lorenz 80 (L80) model \cite{Lorenz80} following \cite[Sec.~3.4]{CLM19_closure} and \cite{chekroun2021stochastic}. {\bf Panel A:} The OPM is the BE manifold shown by blue dots. It provides the slow motion of the L80 dynamics. The L80 dynamics (black curve) evolves onto this manifold and experiences excursions off this manifold, corresponding to bursts of fast oscillations  caused by IGWs (see {\bf Box 3}).  The residual off the BE manifold is 
mainly orthogonal to it causing memory terms to be negligible, and making their stochastic modeling central 
as the $\eta(t)$-term in the MZ decomposition Eq.~\eqref{Eq_GLE}. {\bf Panel B}:  {\mk This residual in the time-domain is  strongly correlated in time and can be grouped in pairs that are} narrowband in frequency and modulated in amplitude with {\mk possible combination of `tones' (bottom panel)}.  {\bf Panel C}:  {\mk Networks of stochastic oscillators such as given in \cite[Eq.~12]{chekroun2021stochastic} are well suited to model such properties}.}\label{Fig_OPMSLO}
\vspace{-2ex}
\end{figure*}

\vspace{-.25cm}
\subsection{The atmospheric Lorenz 1980 model: Markovian and noise terms but no memory}\label{Sec_L80}
\vspace{-.25cm}

Atmospheric and oceanic flows constrained by Earth's rotation
   satisfy an approximately geostrophic momentum balance on larger
   scales, associated with slow evolution on time scales of days, but
   they also exhibit fast inertia-gravity wave oscillations.  The
   problems of identifying the slow component (e.g., for weather
   forecast initialization \cite{bolin1955numerical,baer1977complete,machenhauer1977dynamics,daley1981normal}) and of characterizing slow-fast
   interactions are central to geophysical fluid dynamics. The former was first coined as a slow manifold problem by Leith \cite{leith1980nonlinear}.  
   The Lorenz 63  model \cite{lorenz1963deterministic} famous for its chaotic strange attractor is a paradigm for the
   geostrophic component, while the Lorenz 80 (L80) model \cite{Lorenz80}  is its paradigmatic successor both
   for the generalization of slow balance and for slow-fast coupling. 
   

  \vspace{-.5ex}
  \begin{bclogo}{\small Box 3: The L80 model and bursts of inertia-gravity waves}
  \footnotesize
   The L80 model, obtained as a nine-dimensional truncation of the PE onto three Fourier modes with low wavenumbers \cite{Lorenz80}, can be written as \cite{CLM16_Lorenz9D}:
\beas \label{Eq_L9D}
 a_i \frac{\d x_i}{\d t} &= \e a_i b_ix_jx_k -  c(a_i - a_k) x_j y_k - c^2y_jy_k\\
& \hspace{-1ex}+  c(a_i - a_j) y_j x_k  -  N_0 a_i^2 x_i + \e^{-2}a_i(y_i - z_i), \\
a_i \frac{\d y_i}{\d t} &=  -  \e a_k b_k x_jy_k - \e a_jb_j y_jx_k 
\hspace{-.2ex}+ \hspace{-.2ex}c(a_k-a_j)y_jy_k\\
& \quad -a_i x_i-N_0a_i^2y_i, \\
\frac{\d x_i}{\d t} &=  - \e b_kx_j(z_k-H_k) -\e  b_j(z_j-H_j)x_k \\ 
&+ cy_j(z_k-H_k)  - c(z_j-H_j)y_k + g_0 a_ix_i\\
&\qquad -K_0a_iz_i + \F_i. 
\eeas

The variables $(\x,\y,\z)$ are amplitudes for the divergent velocity potential, streamfunction, and dynamic height, respectively.
Transitions to chaos occurs as the Rossby number $\e$ is increased; see \cite{Gent_McWilliams82,CLM16_Lorenz9D}. 

\begin{wrapfigure}[9]{R}{0.55\textwidth}
\vspace{-3ex}
\centering
\includegraphics[width=.55\textwidth, height=.35\textwidth]{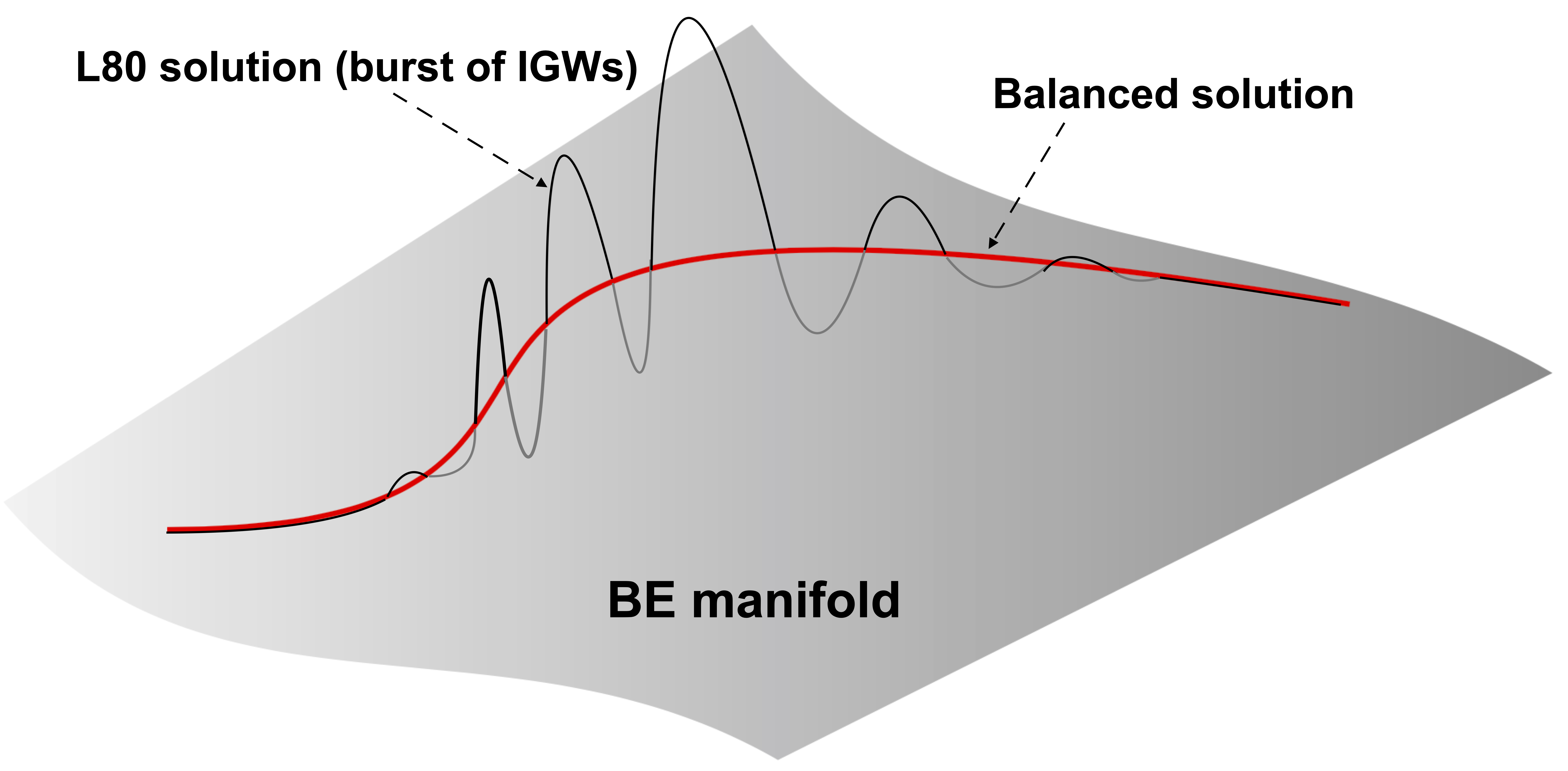} 
\caption*{}\label{BE_visual}
\end{wrapfigure}

At small $\e$, the solutions to the L80 model remain entirely slow for all time (i.e.~dominated by Rossby waves) whereas spontaneous emergences of fast oscillations get superimpose to such slow solutions as the Rossby number is further increased; {\mkr see schematic shown here}.  In such regimes, the balance equation (BE) manifold on which lie balanced solutions \cite{mcwilliams1980intermediate,Gent_McWilliams82} is no longer able to encode the dynamics (see schematic), as the L80 dynamics associated inertia gravity waves (IGWs) get transverse to the BE manifold \cite[Sec.~3.4]{CLM19_closure}.
These regimes with energetic bursts  of IGWs  lie beyond the parameter range Lorenz initially explored in  \cite{Lorenz80} (see \cite{mcwilliams2019perspective}) as well as beyond other regimes with exponential smallness of IGW amplitudes  as encountered in the subsequent Lorenz 86 model \cite{lorenz1986existence,lorenz1987nonexistence,vanneste2008exponential} and the full PE \cite{temam2011slow}  at smaller Rossby numbers; see \cite{vanneste2013balance}.    
\end{bclogo}
Contrarily to other slow-fast systems, this physically-based model exhibits regimes with energetic bursts of fast oscillations superimposed on slow ones {\mkr in each variables of the model which} complicate greatly their parameterization \cite{CLM16_Lorenz9D}; see {\bf Box 3}.    
Regimes beyond exponential smallness of the fast oscillations are not only intimate to the L80 model.  They have been observed in other PE models as conspicuously generated by fronts and jets \cite{plougonven2007inertia,polichtchouk2020spontaneous}, and 
in cloud-resolving models in which large-scale convectively coupled gravity waves spontaneously develop  \cite{tulich2007vertical}. Regions of organized convective activity in the tropics generates also gravity waves leading to a spectrum that contains notable contributions from horizontal wavelengths of 10 km through to scales beyond 1000 km \cite{lane2015gravity} and such IGWs have been also identified from satellite observation of continental shallow convective cumulus  forming organized mesoscale patterns over forests and vegetated areas \cite{Dror2021}.

The L80 model provides a remarkable metaphor of such regimes with  a lack of timescale separation at large Rossby numbers, in which the solutions have slow and fast components (mixture of high and low frequencies (HLF)), 
causing a breakdown of slaving relationships {\mk where the fast variables at time $t$ are a function of the slow variables at the same time instant}, calling thus for a revision of slow manifold methods \cite{leith1980nonlinear} and the like. Only recently, the generic elements for solving such hard closure problems 
have been identified \cite{chekroun2021stochastic}.  Key to its solution is the Balance Equation (BE) manifold  \cite{mcwilliams1980intermediate,Gent_McWilliams82} as rooted in the works of Monin \cite{Monin1952}, Charney and Bolin \cite{charney1955use,bolin1955numerical} and Lorenz \cite{lorenz1960energy}.  The BE manifold has been shown to provide, even for large Rossby number, the slow trend motion of HLF solutions to the L80 model as it  optimally averages out the fast oscillations; nearing this way the OPM, $\Phi_{opt}$, to a high precision \cite{CLM16_Lorenz9D}.

For such regimes, the L80 dynamics evolves onto this manifold and experiences excursions off this manifold, corresponding to bursts of fast oscillations caused by IGWs; see {\bf Box 3} and Fig.~\ref{Fig_OPMSLO}A.  The residual off the BE manifold is 
mainly orthogonal to it,  causing the memory terms  to be negligible \cite{chekroun2021stochastic} and making the stochastic modeling of the $\eta(t)$-term central 
 in the MZ decomposition Eq.~\eqref{Eq_GLE}. An inspection of this residual in the time-domain shows that it is  strongly correlated in time, narrowband in frequency and modulated in amplitude  (Fig.~\ref{Fig_OPMSLO}B). 
 Recent progresses in characterizing the spectral signature in terms of Ruelle-Pollicott resonances and Koopman eigenvalues (and the like \cite{chekroun2017data,zhen2022eigenvalues}) of such time series  \cite{Chekroun_al_RP2,Tantet_al_Hopf}, 
allow for inferring that such residuals can be efficiently modeled by means of a network of Stuart-Landau oscillators  (SLOs) of the {\mkr form
\bea\label{Eq_SLOs}
\dot{z}_j=(\mu_j+&i \omega_j) z_j - (\alpha_j + i \beta_j) z_j |z_j|^2\\
&\; + \textrm{'coupling terms'} + \textrm{'white noise terms'};
\eea
see} Fig.~\ref{Fig_OPMSLO}C and \cite{chekroun2021stochastic} for more details. 
The BE manifold operates here a remarkable feast: It provides a nonlinear separation of variables allowing for decomposing the mixed HLF dynamics of the L80 model into a slow component captured by the BE, and a fast one modeled by a network of SLOs.

{\mkr For the L80 model, the resulting OPM-SLO closure is written for the streamfunction amplitude (the $\y$-variable, see {\bf Box 3})  and takes the form
\be\label{OPM_SLO}
\dot{\y}=\Pi_{\y} \Big(L \y +B\big(\y+\Phi_{opt}(\y)+\xi_t\big) + \bm{f}\Big),
\ee
in which $B$ (resp.~$\bm{f}$) denotes the bilinear (resp.~forcing) terms from the original L80 model,   $\Pi_{\y} $ denotes the projector onto the resolved variable $\y$,  and the stochastic vector $\xi_t$ is modeled by means of the auxiliary networks of SLOs \eqref{Eq_SLOs}. Remarkably, this network of stochastic oscillators through its nonlinear interactions with the parameterization $\Phi_{opt}(\y)$ of the  slow motion  and the $\y$-variable in \eqref{OPM_SLO} allows for recovering with  great accuracy  the multiscale dynamics of the L80 model along with its complex bursts of fast oscillations caused by IGWs; {\mkr see \cite[Figs.~6 and 7]{chekroun2021stochastic}.} }

In terms of Hasselmann's program, the L80 model is thus rich of teachings.   It shows that an efficient modeling of regimes with a lack of timescale separation characterized by a mixture of intertwined slow and fast motions, requires (i) a good approximation of the OPM capturing the  slow motion, (ii) to go beyond stochastic homogenization and the like \cite{majda2001mathematical} to model the noise; the use of network of SLOs showing a great deal of promises in that respect.

Finally it is worth noting that thinking of the bilinear terms $B$ in Eq.~\eqref{OPM_SLO}  as proceeding from advective terms in the L80 model, one may interpret the nonlinear terms involving $\xi_t$ in the stochastic OPM-SLO closure \eqref{OPM_SLO} as stochastic advective terms.  Other recent approaches have shown the relevance of such terms to derive stochastic formulations of fluid flows as well as for emulating suitably the coarse-grained dynamics \cite{memin2014fluid,holm2015variational,cotter2019numerically,resseguier2017geophysical}. {\mkr In particular,  homogenization theory can be used to rigorously derive effective
slow stochastic particle dynamics for the mean part,  under the assumption of mildly chaotic fast small-scale dynamics \cite{cotter2017stochastic}. Interestingly, the route taken for deriving \eqref{OPM_SLO} differs from homogenization techniques, and supports due to the L80 dynamics specificities   that even for regimes in which the timescale separation is violated, closure involving stochastic advection terms may be still relevant.}  

From a practical viewpoint, the interest of disposing of an accurate stochastic closure {\mkr of the stochastic advective form} Eq.~\eqref{OPM_SLO} lies in its ability of  simulating key feature aspects of the multiscale dynamics, offline, in an uncoupled way {\mkr to mimic  the effects of the IGWs}. {\mkr As a result, by simply running offline the network of SLOs \eqref{Eq_SLOs} offline and plugging its  output into Eq.~\eqref{OPM_SLO} as a random forcing input, one recovers by integrating online Eq.~\eqref{OPM_SLO} the multiscale nature of the L80 dynamics through interactions with the nonlinear terms;  see \cite[Figs.~6 and 7]{chekroun2021stochastic}}.  
 
The OPM-SLO approach is thus promising to be further applied to the closure of other more complex slow-fast systems, in strongly coupled regimes.   In particular, regimes exhibiting a mixture of fast oscillations superimposed on slower timescales such as displayed by the L80 model provide a challenging ground for closure in more sophisticated fluid problems.  Such regimes are known to arise in multilayer shallow water models; see e.g.~\cite[Fig.~5]{simonnet2003low}.
In certain regions of the oceans, it has been shown  that IGWs can account for roughly half of the near-surface kinetic energy at scales between 10 and 40 km  \cite{rocha2016mesoscale}, making IGWs energetic on surprisingly large scales. Thus, geophysical kinetic energy spectra can exhibit a band of wavenumbers within which waves and turbulence are equally energetic  \cite{young2021inertia}. We believe in the ability of the OPM-SLO approach to show closure skills for such problems. There, the approximation of the OPM/conditional expectation should benefit from recent progresses accomplished  in neural turbulent closures, as explained below, and the fast component of the motion should also benefit from the wealth of dynamics that networks of SLOs can embody (see Sec.~\ref{Sec_too_manyfreq} below).

\subsection{Neural turbulent closures: No memory, no noise, but spatially non-local Markovian terms}\label{Sec_ML_turb}

Much efforts have been devoted lately into the learning of successful neural parameterizations for the closure of fluid models in turbulent regimes such as the forced Navier-Stokes equations or quasi-geostrophic flow models on a $\beta$-plane; see e.g.~\cite{bolton2019applications,maulik2019,kochkov2021,zanna2020,subel2022explaining}.

These neural closure results are typically obtained with convolutional neural networks (CNNs) \cite{goodfellow2016} that are by definition non-local in space and aim at parameterizing the sub-grid scale stress (SGS) tensor in terms of coarse-grained variables. Among the achievements accomplished by these neural closures, have been reported their ability to provide accurate closures for cutoffs within the
inertial range and for high Reynolds numbers, outperforming more standard schemes such as based on the Smagorinsky parameterizations and the like.

This problem is known to be
difficult as small errors at the level of the SGS typically amplify the errors
at the large scales due to the inverse cascade
\citep{piomelli1991subgrid,jansen2014parameterizing}. To dispose of SGS
parameterizations  at low cutoff levels for such turbulent flows with a
controlled error is thus one of the challenges to resolve.  The accuracy and
stability of the closure results in \cite{bolton2019applications,maulik2019,kochkov2021,zanna2020,subel2022explaining}. are thus strongly supportive for the existence
of a nonlinear function $\boldsymbol{\tau}_{CNN}$ such that the SGS, $\boldsymbol{\tau}$, satisfies, after
spin up, a relation of the form
\beq\label{Eq_fundamental_relation}
\boldsymbol{\tau}=\boldsymbol{\tau}_{CNN}(\overline{u},\overline{v})+\epsilon,
\eeq
where the residual $\epsilon$ is a spatio-temporal function whose fluctuations
are controlled and small in a mean square sense, while $\overline{u}$ and $\overline{v}$ denote the coarse-grained velocity variables {\mkr (not to be confused with $\u$ and $\vb$ used in Sec.~\ref{modelreduction})}. 
Actually, \eqref{Eq_fundamental_relation} is a consequence  of the very construction of $\Phi_{CNN}$ obtained by minimization of loss functions of the form \eqref{Min_tau_int} up to some regularization term.

In
Eq.~\eqref{Eq_fundamental_relation}, $\boldsymbol{\tau}_{CNN}$ denotes the function found by
means of shallow CNNs trained by minimizing a loss function reminiscent to that involved in \eqref{Eq_min_memory}. The
relation \eqref{Eq_fundamental_relation} based on the quality of the closure 
results reported in \cite{bolton2019applications,maulik2019,kochkov2021,zanna2020,subel2022explaining} suggests thus that $\boldsymbol{\tau}_{CNN}$, in the respective cases, is close to the conditional expectation $\overline{\boldsymbol{\tau}}$ \citep{CLM19_closure}, namely the best
nonlinear functional averaging out the unresolved variables as conditioned on
the coarse variables. Thus, {\mkd based on these results, it seems that} finding a good approximation of $\overline{\boldsymbol{\tau}}$ is
sufficient for the {\mkd accurate} closure of forced two-dimensional turbulence problems at
high $Re$, {\mkd at least for a range of physically and computationally interesting cutoffs within the inertial range}.

As such, these neural closure results {\mkdd seem to rule out for such turbulent flows and choice of cutoffs,  the use of memory  terms in the MZ expansion (Sec.~\ref{Sec_MZ}), questioning thus the need of memory terms in other closure studies for similar problems; see
e.g.~\citep{miyanawala2017efficient,parish2017non,ma2019model}. For instance,} memory terms have been advocated for the closure of Kuramoto-Sivashinsky (KS) turbulence in the reduced state space spanned by the unstable modes in e.g.~\cite{lu2017data,ma2019model}, whereas {\mkdd the learning of the} conditional expectation has been shown to be amply sufficient for  {\mkdd high skill closure retaining only the unstable modes and} for even more turbulent regimes \cite[Sec.~6]{CLM19_closure}.

The neural turbulent closure results of \cite{bolton2019applications,maulik2019,kochkov2021,zanna2020,subel2022explaining} restore thus some credentials to ideas proposed in the late 80s by
\cite{FMT88,foias1991approximate} envisioning two-dimensional turbulence as
essentially finite-dimensional with turbulent solutions lying in some thin
neighborhood, in a mean square sense, of a finite-dimensional manifold \citep[Eq.~(1.5)]{CLM19_closure};  ideas that were  watered down as
shown to be valid, only for cutoff wave numbers within or close to the dissipation range \citep{pascal1992nonlinear} when relying on
{\it traditional analytic parameterizations} such as initially proposed in \citep{FMT88}.
 The usage of neural
networks shed thus new lights on this old problem as pushing the validity of
relationships such as \eqref{Eq_fundamental_relation} for cutoff within the
inertial range. {\mkr It is worth noting though that lowering the latter will inevitably be prone at some point to the emergence of memory and/or stochastic terms. Such is the case for instance in closing KS turbulence when the cutoff scale is chosen such that  unstable modes are present in the space of scales to parameterize.}

  \begin{figure*}
	\centering
		\includegraphics[width=.85\textwidth,height=0.45\textwidth]{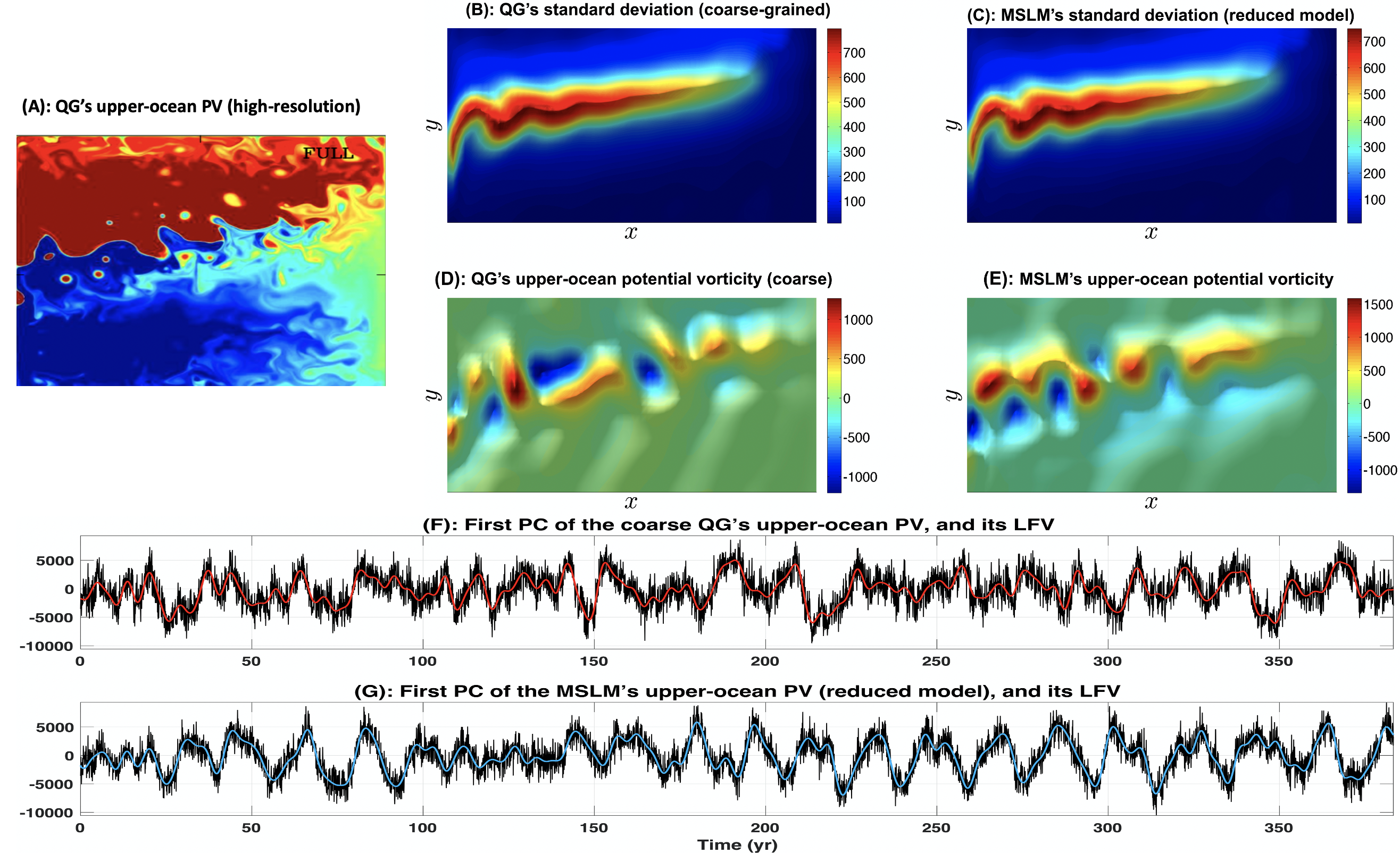}
		\vspace{-1.4ex}
	\caption{\small {\bf Upper-ocean potential vorticity (PV) anomalies, its time-evolution,  
	and standard deviation: QG vs MSLM} (from \cite{Kondrashov_al2018_QG}). {\bf Panel (A)}:  Instantaneous upper-ocean PV anomaly field from a high-resolution simulation (HRS) of a baroclinic QG turbulent model (from \cite{Berloff2016}).. {\bf Panel (B)}: Standard deviation of this HRS over a $64\times 26$ coarse-grid. 
{\bf Panel (C)}:  Standard deviation as simulated by the MSLM reduced model over the same coarse grid. 
	{\bf Panels (D) and (E)}:  Instantaneous upper-ocean PV anomalies  from the coarse-grained QG model and its MSLM reduced model, respectively.
{\bf Panel (F) and (G)}: First PC of the coarse QG's (resp.~the MSLM) upper-ocean PV, and its decadal LFV content shown in red (resp.~blue). The MSLM is able to remarkably reproduce the multiscale temporal variability of the QG coarse-grained dynamics.}
	\label{QG_FLD}
	\vspace{-1ex}
\end{figure*}

\subsection{The {\mkr choice of latent} variables: A baroclinic ocean model example}\label{Sec_too_manyfreq}
We should not looe sight that the MZ framework is conditioned to the choice of resolved and neglected variables inherent to that of the reduced state space in which a closure is sought. This {\mkr aspect} is actually a key step {\mkr when dealing with the equation discovery problem from time-sequential data that have large spatial dimensions such as in climate applications. Then, one} typically compress the {\mkr original set of variables through dimensionality reduction techniques}, into a few variables {\mkr aimed at simplifying the computational burden of finding the governing equations.} 
{\mkr In modern language, the goal is to reduce the number of features that describe the data where the encoder compresses the data from the initial space to the encoded space, also called latent space,  whereas the decoder decompress them.  

The most common method of dimensionality reduction is the principal component analysis (PCA)  also known as empirical orthogonal function (EOF) decomposition \cite{lorenz_1956,jolliffe2002principal}. There, the so-called principal components (PCs) constitute the latent variables. PCA has been commonly used to infer, out of various climate fields, stochastic differential models  that are either linear \cite{Penland89,PenlandMagorian1993,PenlandGhil_MWR93,PenlandSardeshmukh_JCL95} or include nonlinear terms  and various degrees of approximations of memory terms \cite{Franzke2005,kkg05_enso,Franzke2006,kkg06,ckg11,chen2016diversity}. 
Many other methods of dimensionality reduction could be used at this stage such as those based on nonlinear \cite{scholkopf1998nonlinear,mukhin2015principal} or probabilistic versions \cite{tipping1999probabilistic} of EOFs,  spectral versions of EOFs \cite{schmidt2019spectral} and the like \cite{chekroun2017data,zerenner2021harmonic}, {\mkdd transfer/Koopman operators \cite{das2019delay,Froyland2021}, Laplacian and Diffusion maps \cite{belkin2003laplacian,coifman2006diffusion,giannakis2013nonlinear}}, or variational encoders \cite{kingma2013auto}, to name a few.

Whatever the dimensionality reduction method retained, the latent variables to model may display {\mkdd a frequency mixture issue}  as encountered in Sec.~\ref{Sec_L80} for the L80 model, {\mkdd impacting the timescale one wishes to resolve via a reduced model. This issue is e.g.~encountered when aiming at reduced models on decadal timescales of fully-resolved} wind-driven baroclinic quasi-geostrophic (QG) models of the ocean. The ocean circulation of eddy-resolving simulation with $\sim 10^6$ spatial degrees of freedom at reference model parameters \cite{berloff2015dynamically} is characterized by a robust large-scale decadal low-frequency variability (LFV) with a dominant 17-yr cycle, involving coherent meridional shifts of the eastward jet extension separating the gyres; see Fig.~\ref{QG_FLD}A. To this decadal variability is superimposed an interannual variability caused by the eddy dynamics; see  \cite{Kondrashov_al2018_QG}.
Due to this highly turbulent and multiscale nature of the flow,  the capture of the eddies' dynamics on a coarse-grid by a reduced model is highly challenging.  

Within the reduced state space of (the first few dominant) PCs this challenge is manifested by the multiscale nature of the PCs' temporal evolution; a slow evolution (decadal) contaminated by ``fast'' interannual oscillations (due to the eddy-dynamics); see {\mk Fig.~\ref{QG_FLD}F}.  
Such multiscale features constitute the main cause behind e.g.~the failure of multilayered SDEs such as those of \cite{kravtsov2005multilevel,Majda_Harlim2012}  in approximating, here,  the memory and noise terms in the MZ-decomposition, in spite of their successes  in other geophysical problems  as recalled in Sec.~\ref{Sec_variational}.   
The reason behind lies in the set of predictor functions used for the learning of the multilayered SDEs ingredients, either responsible for an explanatory deficit, or {\mkr subject to} a spectral bias if {\mkr convontional} neural networks {\mkr such as multilayer perceptrons} are employed \cite{rahaman2019spectral}.  

This is where, multivariate signal decomposition methods such as  \cite{GhilEtAl_RG02,chekroun2017data} {\mkdd may offer an} alternative {\mkdd route for overcoming these difficulties}, {\mk by extracting {\mkdd from data, frequency-ranked coherent} modes of variability}. Indeed, such methods, when effective in separating the slow and fast temporal components of the PCs (or analogues), provide a natural ground for the modeling of these temporal components by means of stochastic SLOs such as in Eq.~\eqref{Eq_SLOs}, this time ranked by frequency to be resolved, and introduced as Multiscale Stuart-Landau Models (MLSMs) in \cite{chekroun2017data}.
{\mkdd MSLMs have demonstrated skills in modeling challenging Arctic sea ice datasets with nonlinear trends \cite{KCYG_2018_arctic,Kondrashov2018_arcticextent}.
Here, as summarized in Fig.~\ref{QG_FLD} the MSLMs provide coarse-grained models with high closure skill for QG turbulent flows; see  \cite{Kondrashov_al2018_QG} for more details. The latter results} invite for more studies exploiting MSLMs and signal decomposition methods to tackle the closure of more realistic PE models, as well as for more understanding.  The MSLMs being {\mkdd made of} stochastic oscillators, it raises  furthermore the question whether the original quasiperiodic Landau's view of turbulence \cite{landau2013fluid}, despite having been dethroned by the Ruelle and Takens vision based on chaos {\mkd theory} \cite{ruelle1971nature}, may be in the end well suited to describe turbulent motions with the amendment of the inclusion of stochasticity.

}

\section{Describing the Climate Crisis via Response Theory}\label{climateresponse}
We can address the problem of quantifying the  climatic response to forcings by considering the impact of perturbations on the statistical properties of ensemble of trajectories evolving according to a given SDE. Hence, we {\color{black} take Hasselmann's ansatz for a stochastic climate model and}  consider the following $d$-dimensional It\^o SDE 
\begin{equation} \begin{aligned}\label{eq:sto ode 2}
\d \x  = &\left[\FF (\x)  + \sum_{u=1}^U\varepsilon^u_1 g^u_1(t)\GG_u(\x) \right]\d t\\
&+\left[\Sigmab(\x)+\sum_{v=1}^V\varepsilon^v_2g^v_2(t)\Gamma_v(\x)\right]\d \W, 
\end{aligned}\end{equation}
where
the unperturbed dynamics is  given by Eq.~\eqref{Hass1}. {\color{black}Hence, here $\x$ indicates a vector of slow climatic variables.}
We consider the case of general time-dependent perturbations acting on either the deterministic component (a parametric modulation of the system) or in the stochastic component (a perturbation to the noise law) associated e.g.~with changes in the properties of the unresolved degrees of freedom. The perturbations to the drift term are  embodied by the vector fields $\GG_u$, each modulated by a {\mk (scalar) amplitude function} $g^u_1(t)$ and a small parameter $\varepsilon_u$. The perturbations to the noise term are embodied by the $d\times p$ matrices $\Gamma_v$, {\mk whose amplitude are controlled by the functions $g^v_2(t)$ and the small parameters $\varepsilon_v$.}

If one considers a background deterministic dynamics {\mk ($\Sigmab(\x)=0$) and the time-dependent forcing
in the drift terms are non-vanishing}, finding the solution $\rho_\varepsilon(\x,t)$ of the FPE corresponding to Eq.~\eqref{eq:sto ode 2} amounts to studying the  properties of the {\mk statistical equilibrium supported by the system's} pullback attractor \cite{Chekroun2011,CLR13,TelJSP,PieriniJSP}. {\color{black}In practical terms, such a measure can be constructed by initializing an ensemble in the infinitely distant past and letting it evolve according to the time-dependent dynamics \cite{Lucarini2017}.} 
Following \cite{Santos2022}, let us assume that $\rho_\varepsilon(\x,t)$ 
can be written as:
\bes
\rho_\varepsilon(\x,t) = \rho_0(\x) + \sum_{u=1}^U\varepsilon^u_1  \rho^u_{1,d}(\x,t) + \sum_{v=1}^V\varepsilon^v_2 \rho^v_{1,s}(\x,t) + h.o.t.
\ees
Such an asymptotic expansion is the starting point of virtually {\mk all} linear response formulas for statistical mechanical systems; see \cite{Lucarini2016,SantosJSP} for a discussion of the radius of convergence of the expansion above.

The expected value of 
$\Psi$ at time $t$ is
\bes
	\langle \Psi \rangle_{\rho_\varepsilon^t}=\int  \d  \rho_\varepsilon(\x,t) \Psi(\x) =  \langle \Psi \rangle_0+  \delta^{(1)}[\Psi] (t)  +  h.o.t.
\ees
where 
the \emph{linear response} is: 
\bea\label{eq:linear response time dependent}
\delta^{(1)}[\Psi] (t) = \sum_{u=1}^U\varepsilon^u_1 &  \left(g^u_1 \bullet \GGG^u_{d,\Psi} \right) (t) \\
&\qquad+ \sum_{v=1}^V\varepsilon^v_2  \left( g^v_2 \bullet \GGG^v_{s,\Psi} \right) (t)
\eea
where``$\bullet$'' indicates the convolution product between the {\mk forcing amplitudes} $g^{u/v}_{1/2}$ and the Green functions $\GGG_{d/s,\Psi}^{u/v}$;  see \cite[Eq.~(8)]{Santos2022} {\mk and {\bf Box 4} for the latter}. {\color{black}In what follows we consider, without loss of generality, observables having vanishing expectation value in the unperturbed state. This can  be achieved by redefining $\Psi$ {\mkr as} $\Psi-\langle\Psi\rangle_0$. In physical terms, this amounts to considering anomalies with respect to the reference climatology.}

As discussed in \cite{Pedram1}, having  {\color{black}explicit} formulas for the linear response of a climate model to perturbation would entail having an exact theory for determining {\mk in particular} the eddy-mean flow feedback. This {\mk is clearly not an easy task as} it would require a fully coherent theory of climate dynamics, which is still far from having been achieved. Hence, we need to find ways to estimate the response operators. The Green's functions shown in the {\bf Box 4} {\mk (Eq.~\eqref{eq:Green})} can be interpreted as lagged correlations between the  observables $\Phi=\ellL_{1,d/s}^{u/v}(\log \rho_0)$ and $\Psi$. This indicates a generalisation of the classical FDT \cite{kubo1966,abramov2007,pavliotisbook2014}. 
The FDT has been applied in the past to the output of climate models to predict the climate response to changes in the solar irradiance \cite{North1993}, GHGs concentration \cite{Cionni2004,Langen2005} as well as to study the impact of localised heating anomalies \cite{gritsun2007}. 

Nonetheless, the use of gaussian or quasi-gaussian approximations for $\rho_0$, which leads to using Green-Kubo formulas, leads to potentially large errors in the estimate of the response {\color{black}in a nonequilibrium system. In \cite{Pedram2} one can find a rather detailed analysis of the reasons why classical FDT methods fail in reproducing the response operators: 
features associated with weak modes of natural variability (which are possibly filtered out in data preprocessing targeted for near equilibrium systems) can have an important role in determining the response; see also discussion in \cite{gritsun2017}.} 

A possible way forward  is to estimate the Green's functions for the observable(s) of interest from a set of suitably defined simulations. 
As shown in \cite{ragone2016,Lucarini2017,Lembo2020} for the case of CO$_2$ forcing, it is convenient to 
perform an ensemble of $N$ simulations where the CO$_2$ concentration is instantaneously doubled, and the runs continue until the new steady state is obtained. The Green's functions are estimated by taking the time derivative of the ensemble average of the response of the model to such a forcing, and can then be used for performing  projections of climate response to arbitrary protocols of CO$_2$ increase.  {\color{black}See {\bf Box 4} for clarifications of how this formalism sheds lights {\mkr on} various classical notions of sensitivity for the climate system.}

\begin{bclogo}{\small Box 4: Green's Functions, {\color{black}Equilibrium Climate Sensitivity, and Transient Climate Response}}\label{GF}
\footnotesize{The Green's functions $\GGG_{d/s,\Psi}^{u/v}$ are key tools for computing a {\mk system's linear} response  to perturbations. {\mk They can be written as} 
\bes
\GGG_{d/s,\Psi}^{u/v}(t) =\Theta(t) \hspace{-1ex}\int  \hspace{-1ex}\d \rho_0(\x) e^{t\ellL^{\ast}_0} \Psi(\x) \LLL_{1,d/s}^{\mk u/v}(\log(\rho_0(\x))), \mbox{i.e.}
\ees
\bea\label{eq:Green}
\GGG_{d/s,\Psi}^{u/v}(t) 
 &=\Theta(t)\int \dd \x \rho_0(\x)
 \Psi(\x(t)) \Phi(x)\\
 &=\Theta(t){\mk C_{\Psi,\Phi}(t)},
\eea
where $\Theta(t)$ is the Heaviside distribution which ensures causality  \cite{ruelle2009,Lucarini2017,Lucarini2018JSP}. 
The operators $\ellL_0^\ast$, $\ellL^u_{1,d}$ and $\ellL^v_{1,s}$ are: 
\begin{subequations}
\begin{align}
& \ellL_0^\ast \Psi = \FF \cdot \nabla \Psi  + \frac{1}{2}\Sigma\Sigma^{T}: \nabla ^2  \Psi, \label{L0aststo} \\
& \ellL^u_{1,d} \rho = -  \nabla \cdot\left( \GG_u  \rho\right), \; 1\leq u\leq U,\\ 
& \ellL^v_{1,s}\rho = \frac{1}{2}\nabla ^2  : \lp \lp \Sigma_v\Gamma^{T} + \Gamma\Sigma_v^{T} \rp \rho\rp,\; 1\leq v\leq V.\label{L1asta}
\end{align}
\end{subequations}
In \eqref{L0aststo}, $\ellL_0^\ast$ is the Kolmogorov operator, the dual of the Fokker-Planck operator $\ellL_0$ associated  with the unperturbed SDE given in Eq.~\eqref{Hass1}, while ":" denotes the Hadamard product. 

The sensitivity of the system as measured by the observable $\Psi$ with respect to the forcing encoded by $\GGG_{d/s,\Psi}^{u/v}(t)$ measures the long-term impact of switching on the forcing and keeping it at a constant value, which corresponds  to choosing a constant (unitary  time modulation). Hence, such sensitivity can be written as $S_{d/s,\Psi}^{u/v}=\int_0^\infty \mathrm{d}t\GGG_{d/s,\Psi}^{u/v}(t)$.  

{\color{black}Indeed, such a relationship allows one to write the Equilibrium Climate Sensitivity (ECS), i.e. the long-term globally averaged surface air temperature increase due to a doubling of the CO$_2$ concentration \cite{IPCC13}, in terms of the corresponding Green's function \cite{ragone2016,Lucarini2017}. Response theory also allows to write an explicit formula \cite{ragone2016} linking ECS to the transient climate response (TCR), i.e. the globally averaged surface air temperature increases recorded at the time at which CO$_2$ has doubled as a result of 1\% annual increase rate, i.e. roughly after 70 years \cite{Otto2013}. Intuitively, one has that ECS is larger than the TCR because of the thermal inertia of the climate system, namely the fact that following the forcing due to increased CO$_2$ concentration, the system needs some time to adjust to its final, steady state temperature.}}\end{bclogo}

\begin{figure}
    \centering
   a)\includegraphics[width=0.4\textwidth]{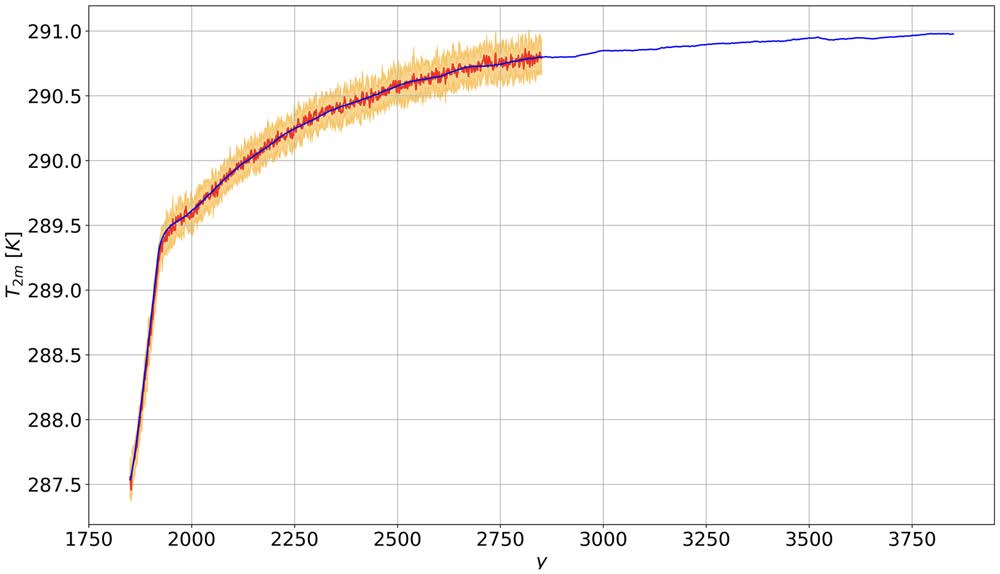}\\
   b)\includegraphics[width=0.4\textwidth,trim=0 5.2cm 0 .2cm,clip]{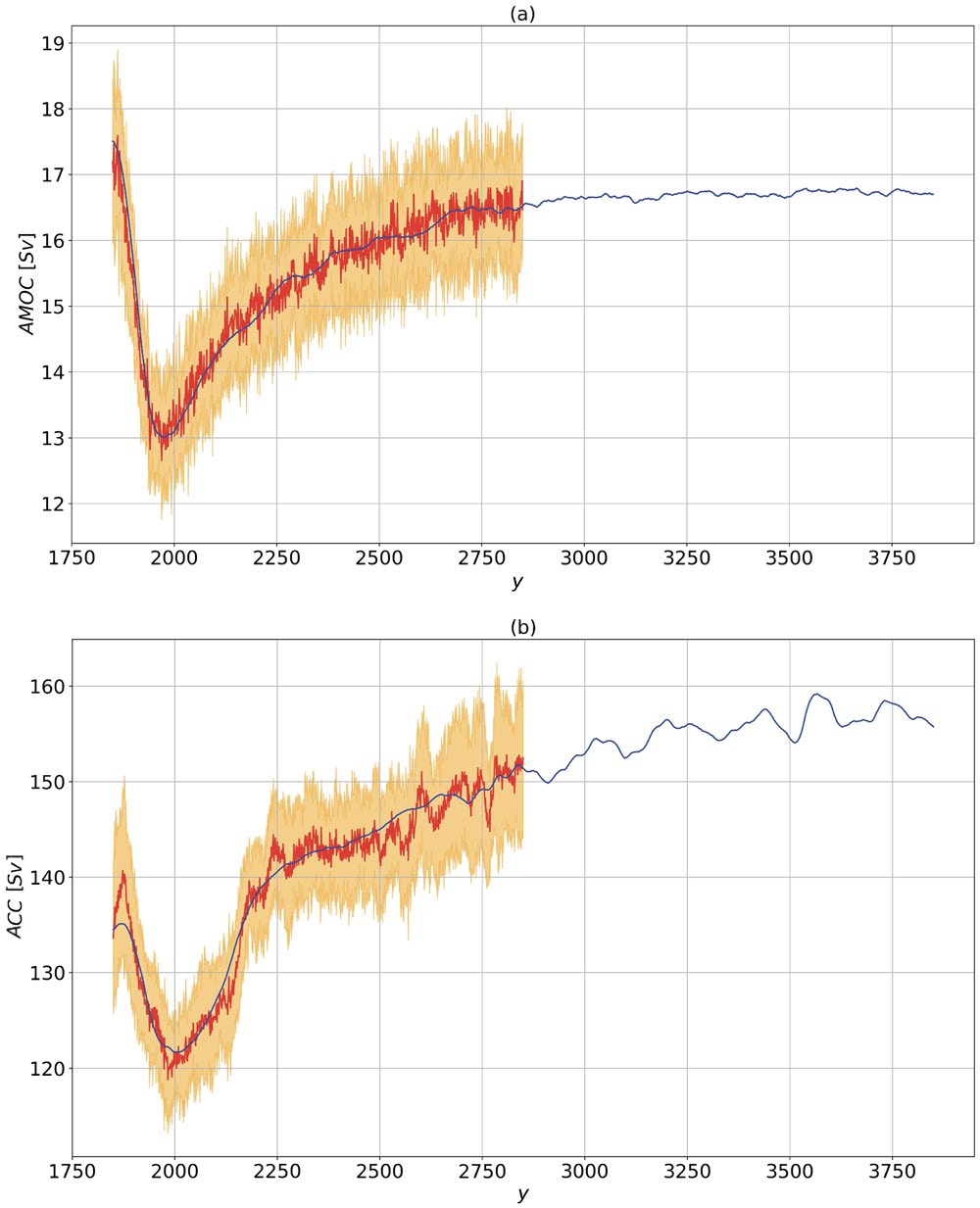}
    \caption{Prediction of a) Globally averaged surface temperature and b) Atlantic Meridional Overturning Circulation (AMOC) strength as a result of an annual 1\% increase of the CO$_2$ concentration from pre-industrial conditions up to doubling. {\mk In each panel} the blue {\mk curve} indicates the {\mk prediction by application of response theory while} the thick red {\mk curve} shows the ensemble mean of the model runs (yellow {\mk curves}).) From \cite{Lembo2020}.}
    \label{fig:Lembo}
\end{figure}

 Figure \ref{fig:Lembo} portrays the application of response theory to an ESM where accurate projections are obtained for the globally averaged surface temperature and for the Atlantic Meridional Overturning Circulation (AMOC) strength \cite{Dijkstra2005low,Kuhlbrodt2007} for a 1\% increase of the CO$_2$ concentration from pre-industrial conditions up to doubling. Note the very pronounced {\mk weakening} of the AMOC and the slow recovery after the applied forcing stabilizes; see discussion in {\mk Sec.~\ref{TPs} below}.

Response theory can be used also for other problems of  practical relevance in climate science, like estimating the point of no return for climate action \cite{Aengenheyster2018} and explaining whether, along the lines of defining causal links, one can use a climate observable as a surrogate forcing acting on another observable of interest \cite{Lucarini2018JSP,Tomasini2020}. 
{\color{black}In \cite{Pedram1} one can find a very useful discussion of additional potential uses of response theory in a climatic context. It can be used to determine the forcing (of given norm) producing the largest response \cite{Antown2018,Antown2022} and to solve {\mkr an} inverse problems like determining the forcing needed to achieve a given response. This viewpoint relates to optimal control ideas \cite[Sec.~3]{CKL17}. A related relevant application pertains the critical appraisal of geoengineering strategies related to the stratospheric injection of aerosols \cite{Bodai2020}.} 

{\color{black}\subsection{Response, Feedbacks, and Koopman Modes}\label{sec:spectral time dependent}}
Response theory makes it possible to investigate the key features of the climatic feedbacks acting on different time scales.  
We provide a rewriting of the Green's functions in terms of individual components specifically associated with the eigenmodes of the unperturbed Kolmogorov operator $\LLL_0^\ast$; see {\color{black}Eq.~\eqref{L0aststo}, {\bf Box 4}}. {\color{black}We assume that $\LLL_0^\ast$ has a vanishing eigenvalue $\lambda_0=0$ with unitary multiplicity, which is tantamount to assuming a unique invariant measure.}   

{\color{black}Let $\{ \lambda_j \}_{j=1}^{M}$ {\mkr denote the} $M$ {\mkr 
nonzero} eigenvalues of finite algebraic multiplicity $m_j$, {\mkr ordered by their real part in decreasing order, i.e.~with $\lambda_1$ having the largest real part}. The $\lambda_j$ are either real or come in complex conjugate pairs. Namely, if $\lambda_j$ is an eigenvalue of the unperturbed Kolmogorov operator $\LLL_0^\ast$ with eigenfunction $\psi^\ast_j$, so is $e^{\lambda_jt}$ of ${K}_t=e^{t\LLL^\ast_0}$ relative to the same eigenfunction. With an abuse of language, we shall refer to the $\lambda_j$'s and the $\psi^\ast_j$ as Koopman eigenvalues and Koopman {\mkr eigenmodes (or simply modes)}, respectively. These {\mkr have} been introduced in Sec.~\ref{Sec_MZ} for the case of deterministic evolution, {\mkr we adopt here the same terminology; the Kolmogorov operator being the analogue to the Koopman generator in the stochastic setting \cite{Chekroun_al_RP2}}.} 
 The eigenfunctions  $\psi^\ast_j$ {\color{black}  encode} the stochastic system's natural variability, decay of correlations and (temporal) power spectra; {\mkr see \cite[Corollary 1 and Sec.~2.3]{Chekroun_al_RP2}.}
Koopman analysis and related methods have demonstrated great promises over the last decade in capturing  modes of climate variability from high-dimensional model and observational data \cite{Froyland2021}.

Following \cite{Tantet2018,Chekroun_al_RP2}, it was shown in \cite{Santos2022} that using the Koopman mode formalism it is possible to express 
the  Green's functions $\mathcal{G}_{k,p}$ introduced in Eq.~\eqref{eq:Green} as a sum of exponential functions (possibly multiplied by polynomials):\be\label{GreenH}
\GGG^k_{d/s,\Psi}(t) =  \Theta(t) \sum_{j=1}^{{M}}\sum_{\ell=0}^{m_j-1} \alpha_j^{\ell,k,s/d}(\Psi)\frac{1}{\ell!} e^{\lambda_jt}t^{\ell},
\ee
where the coefficient $\alpha_j^{\ell,k,s/d}(\Psi)$ are discussed in {\bf Box 5}, whereas their expression can be found in \cite{Santos2022}.
In Eq.~\eqref{GreenH} we are neglecting the contribution to the response coming from the essential component of the spectrum of $\mathcal{L}_0^*$ \cite{Santos2022}. {\color{black}Making this important simplifying assumption amounts, by and large, {\mkr to assuming} that the operator $\mathcal{L}_0^*$ is quasi-compact \cite{engel2000}. This  is} often implicitly assumed when performing {\mk extended} DMD \cite{williams2015data}. This derivation explains why the  formula presented in Eq.~\eqref{GHassExp} allowed {\mkr correctly interpreting} the climate feedbacks in \cite{Maier-Reimer1987,Hasselmann1993b}. We stress here that the $\lambda_j$ do not depend on either the observable or the forcing considered, but are instead a fundamental property of the reference, {\mkr unperturbed} system's dynamics.

{\color{black}\subsection{Tipping Points}\label{TPs}}
Since any Green's function for the observable $\Psi$ can be interpreted as a correlation function between $\Psi$ and a suitably defined observable $\Phi$ ({\bf Box 4}), it should not come to a surprise that, considering Eq.~\eqref{GreenH}, one has:
\be\label{CorrelationH}
{\mk C_{\Psi_1,\Psi_2}(t)}=  \sum_{j=1}^{{M}}\sum_{\ell=0}^{m_j-1} \beta_j^{\ell,k}(\Psi_1,\Psi_2)\frac{1}{\ell!} e^{\lambda_jt}t^{\ell},
\ee
for any pair of observables $\Psi_1,\Psi_2$, as proved in \cite[Corollary 1]{Chekroun_al_RP2}. {\color{black}It is clear from {\mkr Eqns.~\eqref{GreenH}-\eqref{CorrelationH}} that in order to have a non-explosive behaviour, one must have $\mathfrak{Re}(\lambda_j)<0$, $1\leq j\le  M$. From Eqns.~\eqref{GreenH}-\eqref{CorrelationH} it is also clear that the rate of decay of any Green's function {\mkr and that} of any lagged correlation is dominated for large times by the real part of the subdominant pair of the $\mathcal{L}^*_0$ spectrum. We refer to $\gamma=\mathfrak{Re}(\lambda_1)$ as the {\mkr corresponding} \textit{spectral gap}.} The eigenfunctions {\mkr corresponding} to the slowly decaying eigenvalues are the rigorous counterpart of the so-called neutral climate modes \cite{Navarra1993,Palmer1999,Pedram2,Lu2020}.

{\color{black}If we assume that say one  parameter $\eta$ of our system  {\mkr (not to be confused with $\eta$ in Eq.~\eqref{Eq_GLE})} is such that the spectral gap $\gamma=\mathfrak{Re}(\lambda_1)$ vanishes as $\eta\rightarrow\eta_c$, we have that as $\eta$ nears it critical value $\eta_c$}:
\vspace{-.5ex}
\begin{itemize}
\item[(i)] Any Green function  and any lagged correlation
decays sub-exponentially unless the corresponding factors ($\alpha$'s and $\beta$'s, respectively) vanish;
\item[(ii)] {\mk Due to the sensitivity formula} given in {\bf Box 4}, one immediately derives that as we near a tipping point, the sensitivity of the system increases {\mkr in consistency with earlier results linking the smallness of the spectral gap $\gamma$ to sensitivity of the statistics \cite{Chek_al14_RP}.}
\vspace{-.5ex}
\end{itemize}
These two phenomena---critical slowing down and diverging sensitivity---are key manifestations of the proximity of tipping points \cite{Lenton.tip.08}. {\mkr If the} dynamics of the system can be approximated, in a coarse-grained sense, by a Ornstein-Uhlenbeck process {\mkr such as in \cite{PenlandSardeshmukh_JCL95}}, another manifestation of {\mkr getting to close to a} tipping point is  
\begin{itemize}
\item[(iii)] The increase in the {\mkr signal's variance \cite{Held2004,scheffer2012anticipating,Boettner2022}.}
\vspace{-.5ex}
\end{itemize} 
In more general terms, (iii) corresponds to an increased sensitivity of the system to background noise; see \cite{Lucarini2012,Santos2022}. 

\begin{figure}
    \centering
 \includegraphics[width=0.49\textwidth,height=0.4\textwidth]{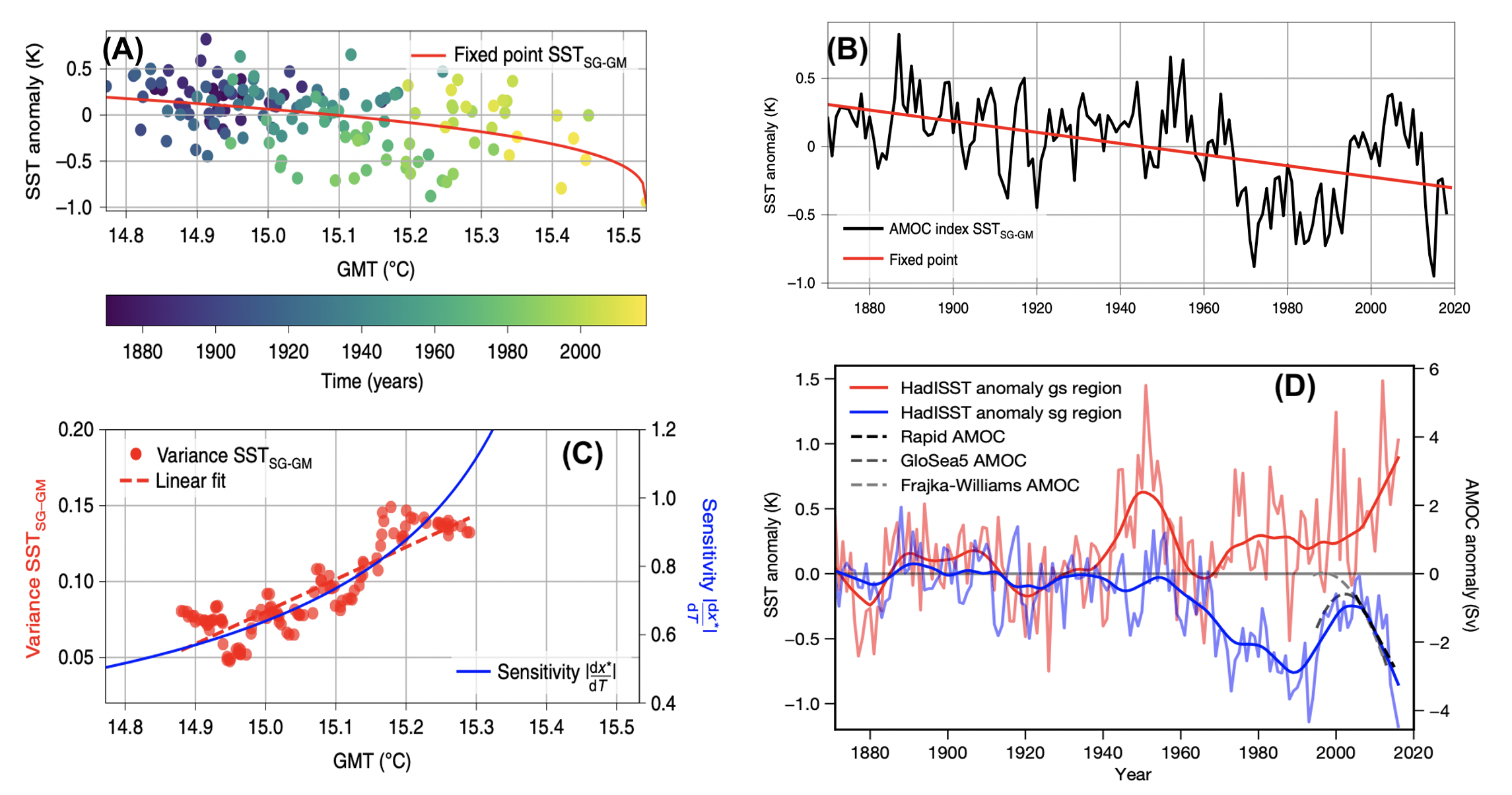}   
 \caption{(Adapted from \cite{Boers2021} and \cite{caesar2018observed}) {\bf Panel A}: AMOC index SST$_{\tiny\mbox{SG-GM}}$ \cite{caesar2018observed}  as a function of global mean temperature (GMT) and least-squares fit of the fixed point of a conceptual AMOC model from \cite{Boers2021}.
 {\bf Panel B}: The SST-based AMOC index SST$_{\tiny \mbox{SG-GM}}$, constructed by subtracting the global mean SSTs from the average SSTs of the subpolar gyre region (black), supplemented by the same least-squares fit (red) \cite{Boers2021}. 
  {\bf Panel C}: Variance of fluctuations of the AMOC index around the fixed point (red) and corresponding sensitivity of the model, with control parameter $T$ given by the global mean SSTs. These variances are estimated over a sliding temporal window and the results are plotted at the centre point of that window \cite{Boers2021}. 
 {\bf Panel D}: Observational evidence of the nearing of the AMOC tipping point during the last century from \cite{caesar2018observed}. Shown are time series of SST anomalies with respect to the global mean SST in the subpolar
gyre (sg) and the Gulf Stream (gs) regions (HadISST data).}
    \label{fig:Boers}
\end{figure}

The AMOC has long been seen as a climatic subsystem with potential tipping behaviour as a result of changing climatic conditions, and, specifically, of changes in the hydrological cycle and {\color{black}cryosphere}  in the Atlantic basin \cite{Rahmstorf1995,Dijkstra2005low,Kuhlbrodt2007}. Figure \ref{fig:Boers} shows {\mk results from \cite{Boers2021,caesar2018observed}} describing {\mk signs from observations that support a nearing of the} AMOC to a {\mk highly plausible} critical transition. {\color{black}One finds 
an increase of the sensitivity and in the variance of an AMOC index tied to the sea surface temperature (SST), as well as by a decrease of the index' rate of decay}; 
{\mk see also \cite[Fig.~4]{Boers2021}}. Large sensitivity and slow recovery of the same {\mk large-scale climate driver has} already been discussed in a modelling context in Fig.~\ref{fig:Lembo}.

We can learn more about such critical behaviour by taking the Fourier transform of the Green{\color{black}'s} function given in Eq.~\eqref{eq:Green}:
\bea\label{eq:susceptibility transform}
\mathfrak{F}\left[\GGG^k_{d/s,\Psi} \right](\omega)=
\sum_{j=1}^{M}\sum_{{\mkr \ell=0}}^{{\mk m_j}-1} \frac{\alpha_j^{\ell,k,s/d}(\Psi)}{\lp i\omega - \lambda_j \rp^{{\mkr \ell+1}}}.
\eea

\begin{bclogo}{{\small Box 5: Nonequilibrium Oscillator Strengths}}
\footnotesize{Equation \eqref{eq:susceptibility transform} indicates that the coefficient $\alpha_j^{\ell,k,s/d}(\Psi)$ weight the contributions to the frequency-dependent response coming from the  eigenmode(s) corresponding to the eigenvalue $\lambda_j$ (which has in general multiplicity $m_j$) for a given combination of observable and forcing. Note that there is a total of $M=M_1+2M_2$ eigenvalues. Of these, $M_1$ are real, and $M_2$ are complex conjugate pairs. Hence, the $\alpha's$ are the non-equilibrium, classical equivalent of the well-known oscillator strengths discussed in spectroscopy, which weight the contributions to the optical susceptibility from each of the allowed quantum transitions from the ground states to the accessible excited states \cite{Hilborn1982,Lucarini2005}. {\mk Thus}, Eq.~\eqref{eq:susceptibility transform} provides the basis for a spectroscopy of general nonequilibrium systems, and, specifically, of the climate system. Resonant terms are associated with tipping phenomena. }
\end{bclogo}

The poles of the susceptibility are located at $\omega= -i\lambda_j$. Equation ~\eqref{eq:susceptibility transform} implies the existence of resonances in the response for real frequencies $\omega=\mathfrak{Im}(\lambda_j)$. Neglecting the possible existence of nonunitary algebraic multiplicities, the susceptibility at the resonance $j$ is proportional to $1/\mathfrak{Re}(\lambda_j)$. Hence, as {\mkr $\eta\rightarrow\eta_c$}, the susceptibility for $\omega=\mathfrak{Im}(\lambda_1)$ diverges, thus implying a breakdown of the response operator. If at criticality $\mathfrak{Im}(\lambda_1)=0$, the static response of the system diverges, indicating a saddle-node-like bifurcation phenomenon {\mk (turning point)}. If, instead, $\mathfrak{Im}(\lambda_1)\neq0$, we face an oscillatory unstable phenomenon that is reminiscent to a Hopf bifurcation \cite{Tantet_al_Hopf,TantetJSPIII}. The viewpoint proposed here allows {\mkr for linking} the {\mk fundamental features} of the tipping phenomenology within a coherent framework.

\section{Detection and Attribution of Climate Change}\label{optimalfingerprinting}
The statistical mechanical tools discussed above allow for performing climate change projections for ensembles of trajectories: statements are made in terms of (changes of) the expectation value of general observables. Clearly this is a mathematical construction that does not fully comply with the requirements of climate science, {\color{black}because, as mentioned above, we wish to make statements on the statistical properties of the climate signal we are experiencing---we {\mkr do} not have to access to the hypothetical multiverse comprising of other statistically equivalent realisations---and possibly link it to acting forcing.  Nonetheless, linear response theory applied to climate provides solid physical and dynamical foundations for detection and attribution {\mkr via} statistical techniques}.

Let $\Psi_k$, $k=1,\ldots,N$ be a collection of climatic variables. {\mk Consider also} the possibility that the forcing to the climate system comes from $F$ different sources, be them anthropogenic or natural. We rewrite  Eq. \eqref{eq:linear response time dependent} by clamping together all the $U+V=F$ acting forcings as $
\delta^{(1)}[\Psi_k] (t) = \sum_{p=1}^F\varepsilon_p  \left(g_{p} \bullet \GGG_{p,k} \right) (t)$. Hence we have that, at first order, 
\begin{equation} \begin{aligned}
\label{eq:da2}
Y_k(t)=\Psi_k(t) -\langle \Psi_k\rangle_0=
\sum_{p=1}^F\tilde{X}_k^p(t) +\mathcal{R}_k(t), 
\end{aligned}\end{equation} 
{\color{black}where $\Psi_k(t)$ indicates the actual value of the climate variable $\Psi_k$ at time $t$,}  
the terms
$\tilde{X}_k^p(t)=\varepsilon_p  \left(g_{p} \bullet \GGG_{p,k} \right) (t)$ 
account for the forced variability, and
$\mathcal{R}_k(t)=\Psi_k(t) -\langle \Psi \rangle_{\rho_\varepsilon^t}$ is a random vector whose correlations are {\mk governed by} the {\mk probability distribution} $\rho_\varepsilon^t$ {\mk solving the FPE associated with Eq.~\eqref{eq:sto ode 2}.} {\color{black}If the variables $\Psi_k$ correspond to anomalies with respect to the reference climatology, we have $\langle \Psi_k\rangle_0=0$.}

Equation \eqref{eq:da2} is 
cast into a form that is very close to the usual mathematical formulation of optimal fingerprinting for climate change given in Eq.~\eqref{eq:da3}. 
Response theory indicates that if we use the forced run of a model to perform detection and attribution of climate change as simulated by the same model, and if we are in the linear regime of response, all the $\beta_p$'s should be unitary, apart from uncertainty. 

We stress that the linear response theory indicates that the optimal fingerprinting procedure could be applied seamlessly for different time horizons of the climate change signal and for suitably linearly filtered signals (e.g. considering time averages).  This suggests that one should perform the optimal fingerprinting for different time horizons at the same time, and check the consistency of the obtained results (in terms of confidence intervals for the $\beta$'s) across the time of the hindcast. Additionally, the fact that linear response theory applies for a large class of forcings and can even be adapted for studying extremes \cite{LKFW14} explains why optimal fingerprinting finds such a broad range of applications{\color{black}, including the analysis of climatic extremes \cite{Naveau2020,Wang2021}.}

{\color{black}Our derivation emphasizes that optimal fingerprinting relies strongly on assuming linearity in the response, with values of the $\beta$-factors different from unity surrogating in a very simplified manner possibly amplified ($\beta>1$) and damped ($\beta<1$) response due to nonlinear effects. A more accurate treatment of nonlinearities could be achieved by adapting the so-called factor separation method, which is used in meteorology to disentangle linear and nonlinear components of the sensitivity of a system to various acting forcings \cite{Stein1993}. Such a method has been recently used for evaluating the factors impacting past climatic conditions \cite{Hossain2023}. The consideration of higher order terms in the response operators \cite{ruelle_nonequilibrium_1998}  could facilitate the development of fingerprinting methods that go beyond the linear approximation. Additionally, it is clear that performing optimal fingerprinting without considering the whole portfolio of acting forcings can lead to misattribution of the causes of detected climate change. The systematic cross-check proposed above for different time horizons might partly take care of these possible criticalities.}   

The term $\mathcal{R}_k(t)$ in Eq.~\eqref{eq:da2} is not associated with the variability of the unperturbed climate---compare with Eq.~\eqref{eq:da3}---but  rather is tied to the system's variability encoded by the probability distribution $\rho_\epsilon^t$; see {\mk also} discussions on time-dependent {\mk probability measures and pullback attractors} in \cite{Chekroun2011,CLR13,CGN17,TelJSP,Tel2020,Chekroun_al22SciAdv}. 
{\mk This allows us to point out} that the classical formulation of optimal fingerprinting lacks {\mk the  framework to account for changes in variability due to} climate change.

The expression for the Green function given in Eq.~\eqref{GreenH} provides useful information for better understanding the robustness of the optimal fingerprinting method. The model error manifests itself in the difference between the spectrum of eigenvalues (and {\mk eigenfunctions}) of the Koopman operator of the ``real'' climate and that of the model used for constructing the fingerprints. Additionally, different climate models will in general feature different Koopman modes and associated eigenvalues. Hence, constructing fingerprints by bundling together information derived from different models seems not so promising in terms of reducing model error.

It is also clear that if only one between the actual climate system and the model used for optimal fingerprinting are close to a tipping point, one expects  major uncertainties and biases because of the qualitative mismatch between the leading $j=1$ term of all the involved Green functions, and, hence, between the model fingerprints and the actual climatic response to the considered forcings. This becomes even more critical if more models are used for estimating of the fingerprints, because  heavily spurious information could be could added. 

\section{Discussion}\label{conclusions}
The Hasselmann program has had a great influence in the modern development of climate science, both regarding its everyday practice and its more foundational aspects. The fundamental idea boils down to treating noise not like a nuisance one needs to filter out to gather useful information, but rather as a key aspect of the climate system one needs to fully explore and appreciate in order to further its understanding and to link model output and observational data. This is key for making progress in detecting and attributing the climate change signals at different spatial and temporal scales. 

The emphasis on the role of noise in creating - somewhat counterintuitively - meaningful signal at lower frequencies has led to fundamental discoveries like in the case of the mechanism of stochastic resonance, which stemmed as a direct application of Hasselmann paradigm in a climatic context \cite{Benzi1981,Nicolis1981,Benzi1982,Nicolis1982} but has since had huge success in a plethora of other research areas \cite{gammaitoni1998}.  Another important area of application of Hasselmann paradigm deals with one of the age-old problems of dynamical meteorology: the  fast atmospheric processes due to baroclinic disturbances have been interpreted as acting as effective noise responsible for the low-frequency variability of the mid-latitudes due with the transitions betweeen competing regimes of circulations, mainly associated with zonal flow and blockings, respectively \cite{Charney1979,Benzi1986,Benzi1989,Kimoto1993a,Itoh.Kimoto.1996}.  

 {\color{black}The (joint) analysis of climate variability and climate response presented here assumes that we are considering forced or free fluctuations around a stable, balanced state, where overall negative feedbacks dominate and control the relaxation processes.  Yet, the very concept of balance might need a critical appraisal when considering ultralong time scales \cite{Arnscheidt2022Balance}, because the internal feedbacks of the system can change sign when different characteristic time scales are considered \cite{Arnscheidt2022Feedbacks}. The prevalence of positive feedbacks indicate processes associated with transitions between possibly radically different climatic regimes \cite{Rothman2017,GhilLucarini2020}.  Also in this case the stochastic climate modelling angle shows great potential. Taking advantage of Fredilin-Wentzell theory of noise-induced escapes from attractors \cite{freidlin1998} and of large deviation theory \cite{T09}, it is possible to develop a theory of metastability for geophysical flows \cite{Bouchet2012,Herbert2015} and for the climate system \cite{LucariniBodai2019PRL,LucariniBodai2020,Margazoglou2021,Rousseau2023}. While gaussian noise is often used in many investigations for reasons of practical and theoretical convenience, more general classes of noise laws---specifically $\alpha$-stable L\'evy processes---are sometimes invoked for studying climatic transitions where sudden jumps between competing states are observed \cite{Ditlevsen1999,penland_levy,Gottwald2021,LucariniSM2022}. For reasons of space and internal coherence, we have chosen not to cover these important research areas in this review.}

The Hasselmann program, by construction, leaves out the problem of finding the root causes of the noise that impacts the slow climatic variables. In this sense, it can be seen as proposing a phenomenological theory of climate. As well known, the noise comes from the fast fluid dynamical instabilities occurring at different scales, in the atmosphere and in the ocean, leading to chaotic {\mkr behaviour \cite{GhilChildress1987,Ghil.2019,GhilLucarini2020}.} The details of the fast processes, in fact, do matter, exactly because there is no time separation one can use to separate the variables of interest from those one wants to parametrize, so that the kind of parsimony one would derive from the use of homogeneization theory cannot be recovered \cite{pavliotisbook2014}. On the other hand, there is no dichotomy between deterministic behaviour and stochastic representation, because chaos generates stochastic processes. In order to advance in the direction of constructing a theory of climate able to account for variability and response to forcing, and able to provide useful and usable information  to address the climate crisis, there is need to inform the stochastic angle on climate with the key details obtained by a  multiscale analysis of the dynamical processes. This review is a preliminary attempt to go in this direction.

\section*{Acknowledgements}
VL acknowledges the support received by the European Union's Horizon 2020 research and innovation program through the projects TiPES (Grant Agreement No. 820970) and CriticalEarth (Grant Agreement No. 956170) and by the EPSRC through the Grant EP/T018178/1. MDC acknowledges the European Research Council  under the European Union's Horizon 2020 research and innovation program (grant no.~810370) and the Ben May Center grant for theoretical and/or computational research. {\mk This work has been also partially supported by the Office of Naval Research (ONR) Multidisciplinary University Research Initiative (MURI) grant N00014-20-1-2023.}
Finally, the authors are grateful to many close collaborators over the years without whom this review would have not been possible: {\color{black}Peter Ashwin}, Pavel Berloff, Richard Blender, Tam\'as Bodai, Niklas Boers, Henk Dijkstra, B\'ereng\`ere Dubrulle, Davide Faranda, Klaus Fraedrich, Vera Melinda Galfi, Giovanni Gallavotti, {\mk Nathan Glatt-Holtz, {\color{black}Georg Gottwald}, Andrey Gritsun, Anna von der Heydt,  Dmitri Kondrashov, Ilan Koren, {\mkr Sergey Kravtsov}, {\color{black}Tobias Kuna}, J\"urgen Kurths, Honghu Liu, Frank Lunkeit, David Neelin, Greg Pavliotis, C\'ecile Penland, Francesco Ragone, Lionel Roques, Jean Roux, Manuel Santos Guti\'errez, {\color{black}Sebastian Schubert}, Eric Simonnet, Antonio Speranza, Kaushik Srinivisan, Alexis Tantet, Tam\'as T\'el, Stephane Vannitsem, Shouhong Wang, Jeroen Wouters, Niccol\`o Zagli, {\mkr Ilya Zaliapin}, with a special gratitude expressed to {\mk Alexandre Chorin}, Michael Ghil, James C.~McWilliams, David Ruelle, and Roger Temam} for their {\mk invaluable} guidance and inspirational works.

\end{document}